\documentclass[reprint, aps,longbibliography, nofootinbib, superscriptaddress]{revtex4-1}

\usepackage{amsmath,amssymb,graphicx}
\usepackage{paralist}

\usepackage{geometry}
\usepackage[normalem]{ulem}

\usepackage{todonotes}
\usepackage{hyperref}

\usepackage{units}

\usepackage{color}

\begin{document}

\title{21-cm observations and warm dark matter models}
\author{A.~Boyarsky}
\affiliation{Lorentz Institute, Leiden University, Niels Bohrweg 2, Leiden, NL-2333 CA, The Netherlands}
\author{D.~Iakubovskyi}
\affiliation{Discovery Center, Niels Bohr Institute, Blegdamsvej 17, DK 2100, Copenhagen, Denmark}
\affiliation{Bogolyubov Institute of Theoretical Physics, Metrologichna 14-b, 03143, Kyiv, Ukraine}
\author{O.~Ruchayskiy}
\affiliation{Discovery Center, Niels Bohr Institute, Blegdamsvej 17, DK 2100, Copenhagen, Denmark}
\author{A.~Rudakovskyi}
\affiliation{Bogolyubov Institute of Theoretical Physics, Metrologichna 14-b, 03143, Kyiv, Ukraine}

\author{W.~Valkenburg}
\affiliation{Institute of Physics, Laboratory for Particle Physics and Cosmology (LPPC), \'Ecole Polytechnique F\'ed\'erale de Lausanne, CH-1015 Lausanne, Switzerland}

\begin{abstract}
Observations of the redshifted 21-cm signal (in absorption or emission) allow us to peek into the epoch of the ``Dark Ages'' and the onset of reionization.
These data can provide a novel way to learn about the nature of dark matter,
in particular about the formation of small-size dark matter halos. 
However, the connection between the formation of structures and the 21-cm signal requires knowledge of a  stellar to total mass relation, an escape fraction of UV photons, and other parameters that describe star formation and radiation  at early times.
This baryonic physics depends on the properties of dark matter and in particular, in warm-dark-matter (WDM) models, star formation may follow a completely different scenario, as compared to the cold-dark-matter case. 
We use the recent measurements by EDGES [J.~D.~Bowman, A.~E.~E. Rogers, R.~A.~Monsalve, T.~J.~Mozdzen, and N.~Mahesh, An absorption profile centred at 78 megahertz in the
sky-averaged spectrum, Nature (London) 555, 67 (2018).] to demonstrate that when taking the above considerations into account, the robust WDM bounds are in fact weaker than those given by the Lyman-$\alpha$ forest method and other structure formation bounds.
In particular, we show that a resonantly produced 7-keV sterile neutrino dark matter model is consistent with these data.
However, a holistic approach to  modeling of the WDM universe holds great potential and may, in the future, make 21-cm data our main tool to learn about DM clustering properties.
\end{abstract}

\maketitle

The hyperfine splitting of the lowest energy level  of the neutral hydrogen atom leads to a cosmic 21-cm signal thanks to the abundance of primordial hydrogen.
The 21-cm signal from the post-reionization Universe has been studied by a number of experiments (e.g., LOFAR~\cite{Patil:2017zqk,Gehlot:18}, GMRT~\cite{Paciga:2013fj}, PAPER~\cite{Ali:2015uua} (see however~\cite{Ali:18}), MWA~\cite{Ewall-Wice:2016bhu}), but the only tentative detection of the 21-cm signal in absorption against the CMB background at $z \sim 16-19$ has recently been claimed by the EDGES experiment~\cite{Bowman:2018yin}.\footnote{Note however that the result is still uncertain, and there are alternative, noncosmological explanations~\cite{Hills:2018vyr,Bradley:2018eev}.}
It is clear that the forthcoming  experiments, such as the staged HERA~\cite{DeBoer:2016tnn} or future SKA~\cite{Koopmans:2015sua,Bull:2018lat} will offer detailed information about the distribution of the 21-cm signal, thus allowing for the full 3D tomography of the signal, offering an unprecedented reach into the early Universe.
This makes the study of the 21-cm signal a promising tool to learn not only about cosmological parameters (see, e.g.~\cite{McQuinn:2005hk,Mao:2008ug,Oyama:2015gma}) but also about different properties of dark matter, including its decays and annihilations~\cite{DAmico:2018sxd,Yang:2018gjd,Clark:2018ghm,Liu:2018uzy,Cheung:2018vww,Mitridate:2018iag}, dark matter-baryon interactions~\cite{Barkana:2018lgd, Fialkov:2018xre, Berlin:2018sjs, Barkana:2018qrx, Fraser:2018acy, Slatyer:2018aqg}, and the formation of gravitationally bound structures~\cite{Madau:1996cs,Furlanetto:2006jb,Zaldarriaga:2003du,Ciardi:2004ru,Pritchard:2011xb,Barkana:2016nyr}.

\begin{figure*}
    \centering
    \includegraphics[width=0.48\textwidth]{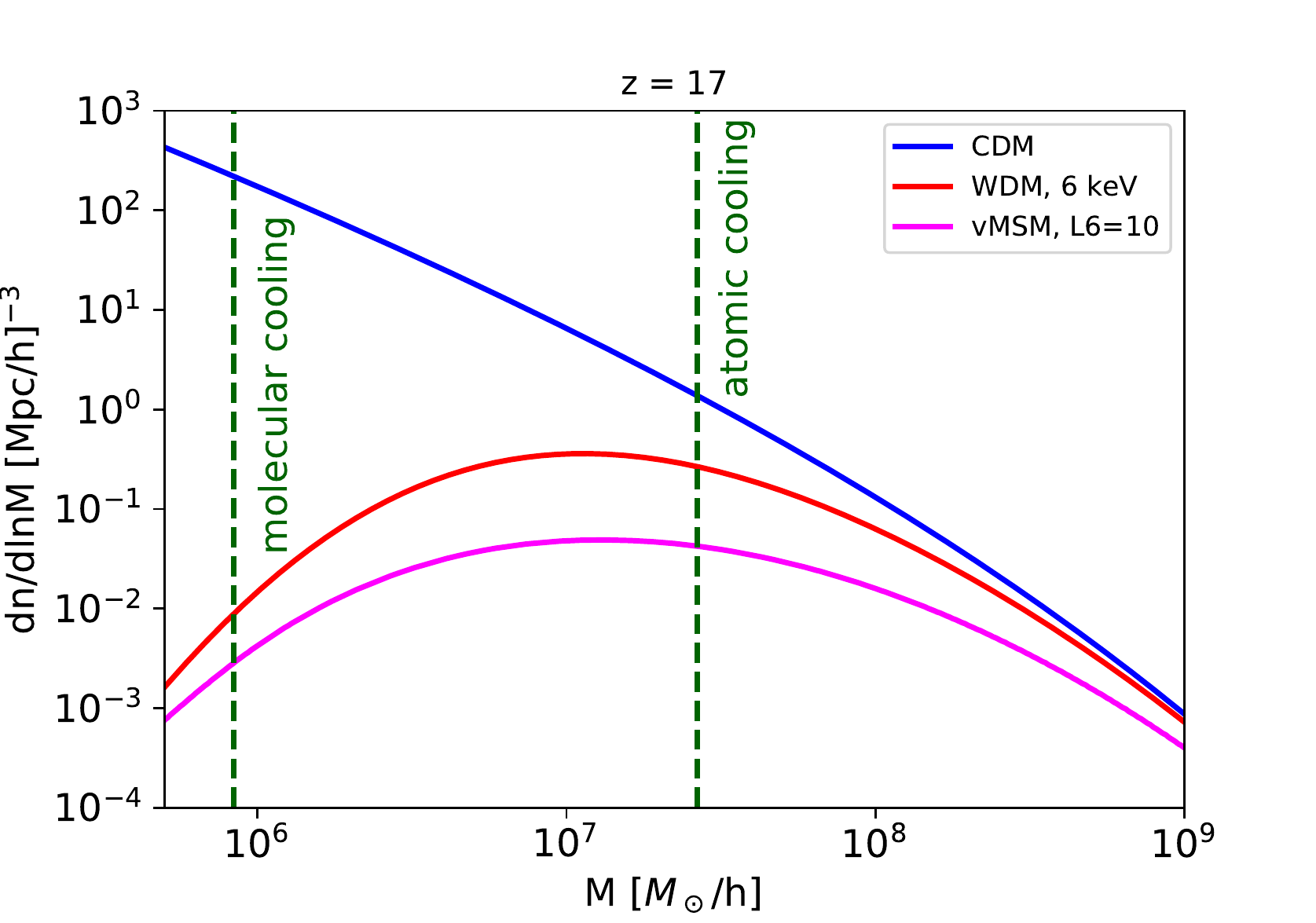}
    \includegraphics[width=0.48\textwidth]{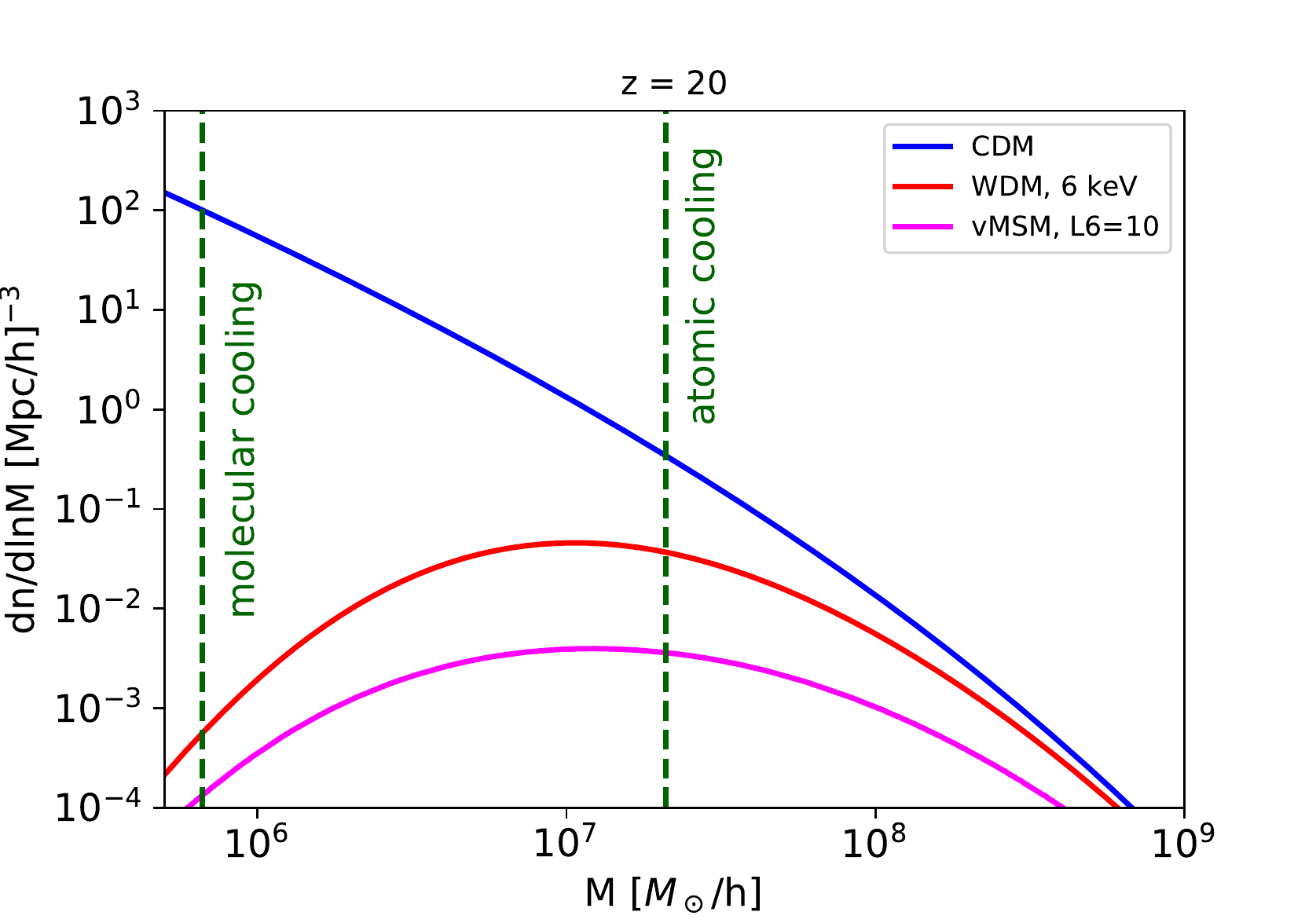}
    \caption{Halo mass functions of models of our interest at redshifts 17 (left) and 20 (right). The masses that correspond to $T_{vir} = 10^3$~K (molecular cooling) and $T_{vir} = 10^4$~K (atomic cooling) are marked as green dashed vertical lines. 
    At both redshifts the molecular cooling threshold has little effect on the collapsed fraction~\protect\eqref{eq:fcoll} in WDM and sterile neutrino models, while for CDM the impact of molecular cooling is substantial, as Fig.~\ref{fig:dTb-009-T1000} illustrates.}
    \label{fig:hmfs}
\end{figure*}

In this work we focus on the global (sky-averaged) 21-cm {absorption} signal that appears when the spin temperature (logarithm of the ratio of population of two levels of the hydrogen's $1S$  state) becomes smaller than the CMB temperature (for a review see, e.g.,~\cite{Furlanetto:2006jb,Pritchard:2008da, Pritchard:2011xb}). 
The standard explanation for this difference of temperatures is the presence of a bath of Ly-$\alpha$ photons which induce transitions between $1S_1$ and $1S_3$ levels:  Ly-$\alpha$ pumping.
Therefore, a detection of the global 21-cm absorption signal at some redshift $z_0$ implies that  sources of radiation  have already been active at that epoch.

With our current knowledge of baryonic physics, we can robustly state that such radiation sources can only form inside dark matter overdensities.
Hence, to predict the 21-cm signal one has to follow several steps:
\begin{compactenum}[\bf a)]
\item Start from the description of bound gravitational structures at a given redshift $z$.
\item Continue with the description of how baryons collapse into  these structures (which depends both on the size of the structures,  on redshift and on cosmology).
\item Assuming a particular type of radiation sources (as they cannot be modeled from first principles), estimate the number of produced photons and model (usually through a combination of semianalytical and numerical methods) how radiation escapes from the bound structures and heats the ambient medium;
\item Given the resulting function of radiation density $d\rho_{\rm rad}/dz$ one can then use available codes (such as ARES~\cite{Mirocha:2014faa} or 21CMFAST~\cite{Mesinger:2010ne}) to  predict the 21-cm signal.
\end{compactenum}

Uncertainties as well as differences in predictions, between DM models are introduced at every step in this process.

\paragraph*{(a) Bound DM structures.}
Historically, the first warm dark matter models were those of sufficiently massive Standard Model neutrinos (see, e.g,~\cite{Bond:1980ha}). Such particles were in thermal equilibrium in the early Universe and froze out while still being relativistic.
They remained relativistic for some period in the radiation dominated epoch and homogenized primordial density perturbations on scales below the free-streaming horizon,  $\lambda_{\rm fs}$ (for a proper definition see, e.g., \cite{Boyarsky:2008xj,Boyarsky:2018tvu}).
The number density of such WDM thermal relics is uniquely determined by the temperature of freeze-out or, equivalently, by their mass, $m_{\rm TH}$.
This mass of the thermal relic is the most typical parametrization of the WDM models.\footnote{An alternative parametrization is given by the mass of non-resonantly sterile neutrinos~\cite{Dodelson:1993je}. The two models lead to an almost identical shape of the matter power spectrum and therefore their masses are related to one another in a nonlinear way; see~\cite{Viel:2005qj,Boyarsky:2008xj} for details. In this work we always indicate what definition of mass we are using. }
All WDM models have suppressed (as compared to CDM) number of halos with masses below the free-streaming cutoff scale, $M_{\rm cut}= \frac{\pi}6 \lambda_{\rm fs}^3$, where $\lambda_{\rm fs}$ is the free-streaming horizon (see, e.g.,\ \cite{Boyarsky:2018tvu}).
This leads to a large difference between a number of collapsed halos, especially at high redshifts, between CDM and WDM models (see Fig.~\ref{fig:hmfs} for our halo mass functions calculated by using the standard prescription proposed in~\cite{Benson:2012su}, also fully consistent with Fig.~1 of~\cite{Schneider:2018xba}).
Naively, one could also expect a big difference between two models in terms of produced starlight. 
However, only the  halos with masses down to $10^7-10^8\, \unit{M_\odot}/h$ contribute to the formation of stars in CDM at redshifts of interest.
Indeed, these masses correspond to virial halo temperatures $\sim 10^3-10^4\, \unit{K}$ -- temperatures that are needed for the hydrogen to cool sufficiently fast, in order to collapse and form compact radiative sources~\cite{Haiman:1999mn,Barkana:2000fd}, see Eq.~\protect\eqref{eq:Mmin} below.

In addition to halos another bound DM structures -- filaments -- can exist in the early Universe. 
Near the cutoff mass formation of filaments and their subsequent fragmentation may be the dominant structure formation process in WDM~\cite{Gao:2007yk,Paduroiu:2015jfa}, as opposed to the CDM model. 
The impact of filaments on the 21-cm signal is  studied by~\cite{Leo:2019gwh} (see also~\cite{Chatterjee:2019jts}), with the outcome that the lower bound on the WDM mass should be weakened  compared with $\gtrsim 6$~keV in earlier works~\cite{Schneider:2018xba,Lopez-Honorez:2018ipk} that did not take into account  this effect. 
In addition to this difference, the presence of filaments also interferes with the structure formation processes, as discussed below.

\paragraph*{(b) Baryonic collapse and star formation in different DM universes}
In general, the naive expectation that what is known from CDM simulations would also apply to WDM universes does not hold up.
Let us point out two remarkable differences between star formation in CDM and WDM.

First, in WDM universes star formation in filaments may dominate over star formation in halos at redshifts $z \gtrsim 6$~\cite{Gao:2007yk,Gao:2014yja}, producing different populations of stars and different amounts of Lyman-$\alpha$ photons.
The star-formation efficiency of these processes is still highly uncertain, but it is clear that they can play a role.
Such a mechanism is absent in CDM.

Second, both hydrodynamical simulations of galaxy formations~(cf. \cite{Herpich:2013yga,Maio:2014qwa,Colin:2014sga,Power:2016ach,Lovell:2016fec}) and semianalytical models (cf.~\cite{Menci:2012kk,Kang:2012up,Menci:2013ght,Nierenberg:2013lqa,Lovell:2015psz,Lovell:2016nkp})
are tuned to reproduce galaxy observables (e.g., luminosity or stellar mass functions, etc.) at $z = 0$.
Not surprisingly this leads to galaxy populations in CDM and WDM having similar properties in recent epochs~\cite{Wang:2016rio}.
However, in order to achieve this agreement one has to choose quite different star-formation prescriptions in CDM and WDM at high redshifts~\cite{Wang:2016rio} , especially for halos close to $M_{\rm fs}$~\cite{Bose:2016hlz}.
As the halo formation in the WDM Universe often starts later, one generically requires higher star-formation efficiencies for WDM (consistent with what we infer in our work).

\paragraph*{(c) Modeling radiation.}
According to the well-developed theory of the 21-cm signal in the early Universe (see, e.g., \cite{Pritchard:2011xb}), the key driver of the timing of 21-cm absorption is the emission rate of Ly-$\alpha$ photons that excite the electrons in hydrogen and result in a spin flip of such electrons after deexcitations (Ly-$\alpha$ pumping). The most common mechanism for emitting Ly-$\alpha$ photons at high redshifts is early star formation~\cite{Pritchard:2011xb} (note however that the QSO contribution can also be significant; see, e.g.,~\cite{Ross:2018uhh}). 

In a CDM universe, the bulk of stars is formed in halos.
Therefore, the star-formation rate density is usually parametrized by the ansatz (see, e.g.,~\cite{Pritchard:2011xb,Mirocha:2014faa,Safarzadeh:2018hhg,Schneider:2018xba,Lopez-Honorez:2018ipk}) $ \dot \rho_{*}(z) = f_{*} \bar \rho_{{\rm b}, 0} \dot f_{\rm coll} (z), $ for redshift $z$, star density (calculated in comoving volume) $\rho_{*}$,  $\dot{\,} \equiv\frac{d}{dt}$ with time $t$, $\bar \rho_{{\rm b}, 0}$ the homogeneous baryon density today, $f_{\rm coll} (z)$ the fraction of baryons in collapsed structures, and $f_*$ the fraction of collapsed baryons that form stars.

The fraction $f_{\rm coll} (z)$ is derived from the halo mass function of a model as 
\begin{equation}
    \label{eq:fcoll}
f_{\rm coll} (z) = \frac{1}{\rho_{m}}\int_{M_{\rm min}}^\infty dM \frac{d\,n}{d\,\ln M},
\end{equation}
with a cutoff for halos below mass $M_{\rm min}$ which are expected not to be able to form stars. This cutoff is set by the halo's virial temperature $T_{vir}$, the temperature which the gas reaches during the virialization of the halo~\citep{Barkana:2000fd}:
\begin{multline}
\label{eq:Mmin}
M_{min} = 1.0\times 10^8 \left(\frac{1+z}{10}\right)^{-3/2}\left(\frac{\mu}{0.6}\right)^{-3/2}\\\times\left(\frac{T_\text{vir}}{1.98\times 10^4~\unit{K}}\right)^{3/2}\left(\frac{\Omega_m}{\Omega_m^z}\frac{\Delta_c}{18\pi^2}\right)^{-1/2}\unit{M_\odot/h},
\end{multline}
where $z$ is the halo redshift, $\mu \simeq 0.60$ is the mean molecular weight, $\Omega_m^z = 1-\Omega_\Lambda/[\Omega_m(1+z)^3+\Omega_\Lambda]$ and $\Delta_c = 18\pi^2 + 82(\Omega_m^z - 1) - 39(\Omega_m^z - 1)^2$~\citep{Bryan:1997dn}. Depending on which mechanism is responsible for cooling, this cutoff may vary: atomic cooling is associated with a cutoff $T_{vir} \simeq 10^4$~K, while molecular cooling leads to a cutoff $T_{vir} \simeq 10^3$~K, see, e.g., Fig.~12 of~\cite{Barkana:2000fd}. The consequences of this parameter are discussed later, and visualized in Fig.~\ref{fig:hmfs}.

Galaxies or galaxy candidates have been observed for $z \lesssim 10$~\cite{Oesch:17}, and we can only extrapolate the aforementioned ansatz for the redshifts of interest.
The star-formation efficiency in halos can be estimated from the observed ultraviolet luminosity function (UV LF) (see, e.g.,~\citep{Dayal:14,Sun:15,Mirocha:16a, Mirocha:2018cih,Park:18}). 
The dependency  $f_*(M,z)$ on halo mass and redshift relies on the model of star formation, and possible values of $f_*$ vary in a wide range. For example, in CDM  halos $f_*$ may reach $0.3$ at $z=5-8$ for $10^{11}-10^{12}\,\unit{M_\odot/h}$ halos, increase with redshifts, and be close to unity during the Dark Ages \citep{Sun:15}. 
In addition the observational estimates of star-formation efficiency depend on assumed cosmology and $f_*$ in low-mass galaxies may be higher in WDM compared to CDM (see, e.g., \cite{Sawala:2014baa, Corasaniti:2016epp, Menci:2018lis}).

Apart from observations, $f_*$ can be predicted in CDM by use of detailed numerical simulations of the Universe during redshifts $z \sim 6-15$~\cite{Sawala:2014baa,Wise:2014vwa,Xu:16,Ma:2017avo,Rosdahl:18,Sharma:2019mrx}. However, there is a three-orders-of-magnitude scatter among the values of $f_*$ in individual simulated galaxies. As Figs 15 and~16 of~\cite{Xu:16} demonstrate, a few galaxies with $f_* \simeq 0.3$ produce an amount of starlight which is several times larger than that of the bulk of galaxies with $f_* \simeq 0.01$. 
As a result, it is \emph{currently impossible} to derive a robust constraint on $\dot \rho_{*}(z \sim 17)$.

An escape fraction of ionizing photons in galaxies during the reionization and Dark Ages has not been determined directly and is  still uncertain (see, e.g., Sec. 7.1 in \cite{Dayal:2018hft}). However, varying the ionizing photon escape fraction in a wide range does not change the redshift of the 21-cm absorption signal significantly.
 The escape fraction of photons in the band $10.2-13.6$\,eV is usually assumed to be close to unity (see Sec.~3.5 of \cite{Mirocha:16a} and references therein).

\paragraph*{(d) Predicting the 21-cm signal}
The above-mentioned uncertainty on $f_*$ translates into a strong systematic uncertainty on WDM parameters that can be probed with a 21-cm absorption signal.
In order to demonstrate this, we computed the 21-cm absorption signal using the ARES code for
three models: CDM, thermal relics with a mass $m_\text{TH} = 6$~keV (claimed to be excluded in~\cite{Schneider:2018xba,Lopez-Honorez:2018ipk}) and the resonantly produced sterile neutrino, with particle mass of 7~keV and lepton asymmetry $L_6 = 10$.\footnote{Lepton asymmetry $L_6 \equiv 10^6 (n_{\nu_e}-n_{\bar{\nu}_e})/s$, where $n_{\nu_e}$ and $n_{\bar{\nu}_e}$ are the number densities of electron neutrinos and antineutrinos, and $s$ is the total entropy density in early Universe~\cite{Laine:2008pg,Boyarsky:2009ix}}
This sterile neutrino model is consistent with all  astrophysical and cosmological bounds: x-ray bounds on decaying DM~\cite{Bulbul:2014sua,Boyarsky:2014jta,Boyarsky:2014ska,Iakubovskyi:2015dna,Ruchayskiy:2015onc,Franse:2016dln,Adhikari:2016bei,Abazajian:2017tcc,Boyarsky:2018tvu}, suppression of the power spectrum as inferred from  the Lyman-$\alpha$ forest~\cite{Garzilli:2015iwa,Baur:2017stq,Garzilli:2018jqh}, cosmic reionization~\cite{Rudakovskiy:2016ngi,Bose:2016hlz,Rudakovskyi:2018jfc}, and Milky Way satellite and galaxy counts~\cite{Lovell:2015psz,Lovell:2016fec}.

\begin{figure*}
    \centering
    \includegraphics[width=0.6\textwidth]{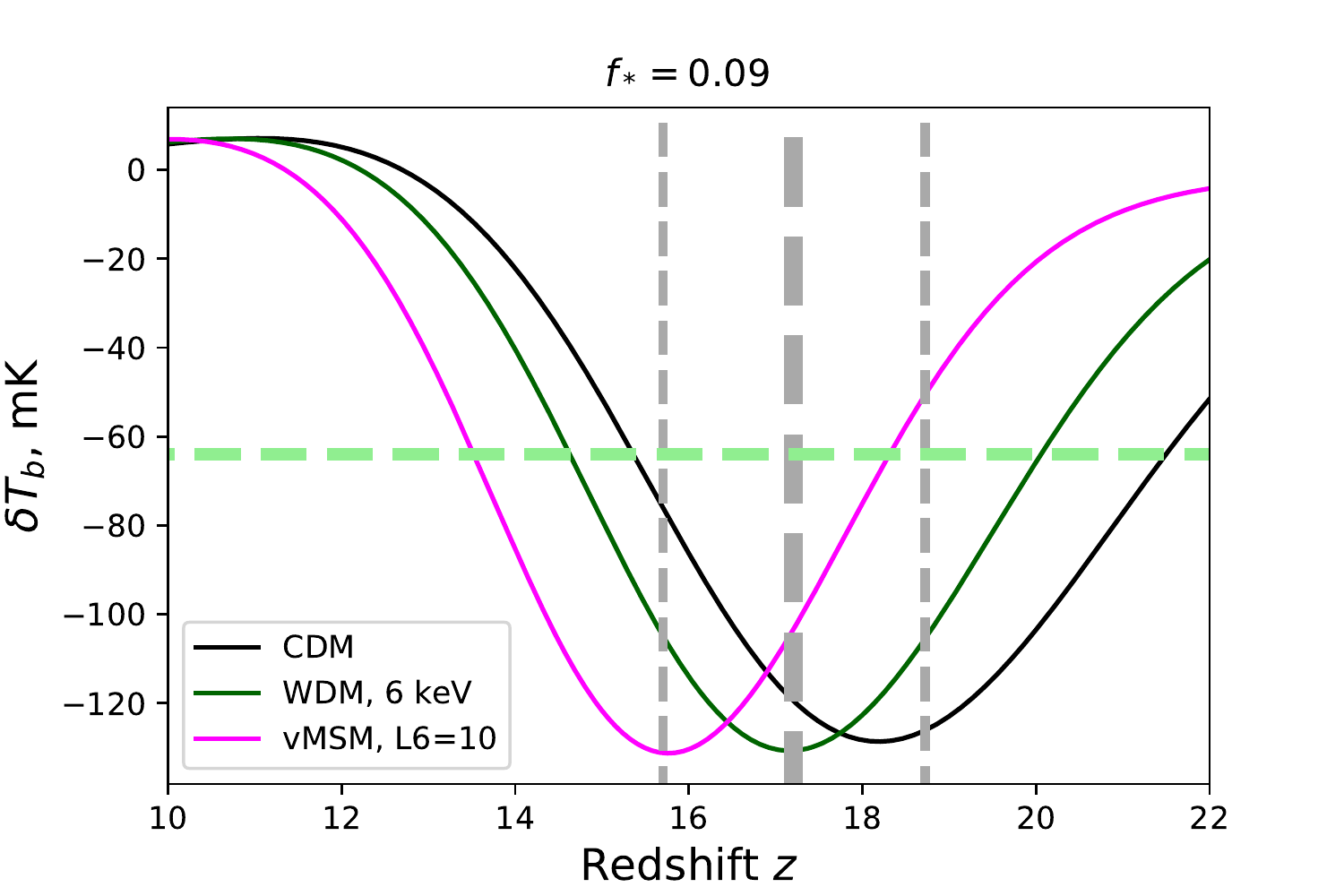}
    \caption{$\delta T_b$ as a function of redshift $z$ for three models of interest: CDM, thermal-relic WDM with mass $m_\text{TH} = 6$~keV, and resonantly produced sterile-neutrino DM with mass $7$~keV and lepton asymmetry $L_6 = 10$. For all models the minimal virial temperature of halos is fixed at $T_{vir} = 10^4$~K, corresponding to atomic hydrogen cooling; see,  e.g., Fig.~12 and Eq.~(26) of~\cite{Barkana:2000fd}. The stellar formation efficiency $f_*$ is chosen to be $0.09$. 
    Due to higher star-formation efficiency as compared to e.g., \cite{Safarzadeh:2018hhg,Schneider:2018xba}, the position of the 21-cm absorption trough becomes consistent with EDGES observations (indicated by the grey vertical lines) for all three models of our interest. The green horizontal line denotes half of the absorption depth; it is plotted in order to illustrate the full width at half maximum of the absorption troughs in the models of our interest.}
    \label{fig:dTb}
\end{figure*}
The results are shown in Fig.~\ref{fig:dTb}. The results strongly depend on the range of assumed values of
 $f_*$.
From the discussion above we see that 
it should be at least from $f_* \simeq 0.01 $ to $f_*\simeq 0.3$ (see, e.g.,~\cite{Xu:16}).
We see that for $f_* = 0.09$ in both 7-keV sterile neutrinos and thermal relics with $m_\text{TR} = 6$~keV, the minimum of $\delta T_b(z)$ happens around $z=17$, in agreement with the EDGES results. 
On the contrary, taking $f_* = 0.03$ (as done in~\cite{Schneider:2018xba}) would make CDM consistent with the EDGES data, while the two WDM models would have an insufficient number of Lyman-$\alpha$ photons at the redshifts of interest.

\begin{figure}[!t]
    \centering
    \includegraphics[width=\linewidth]{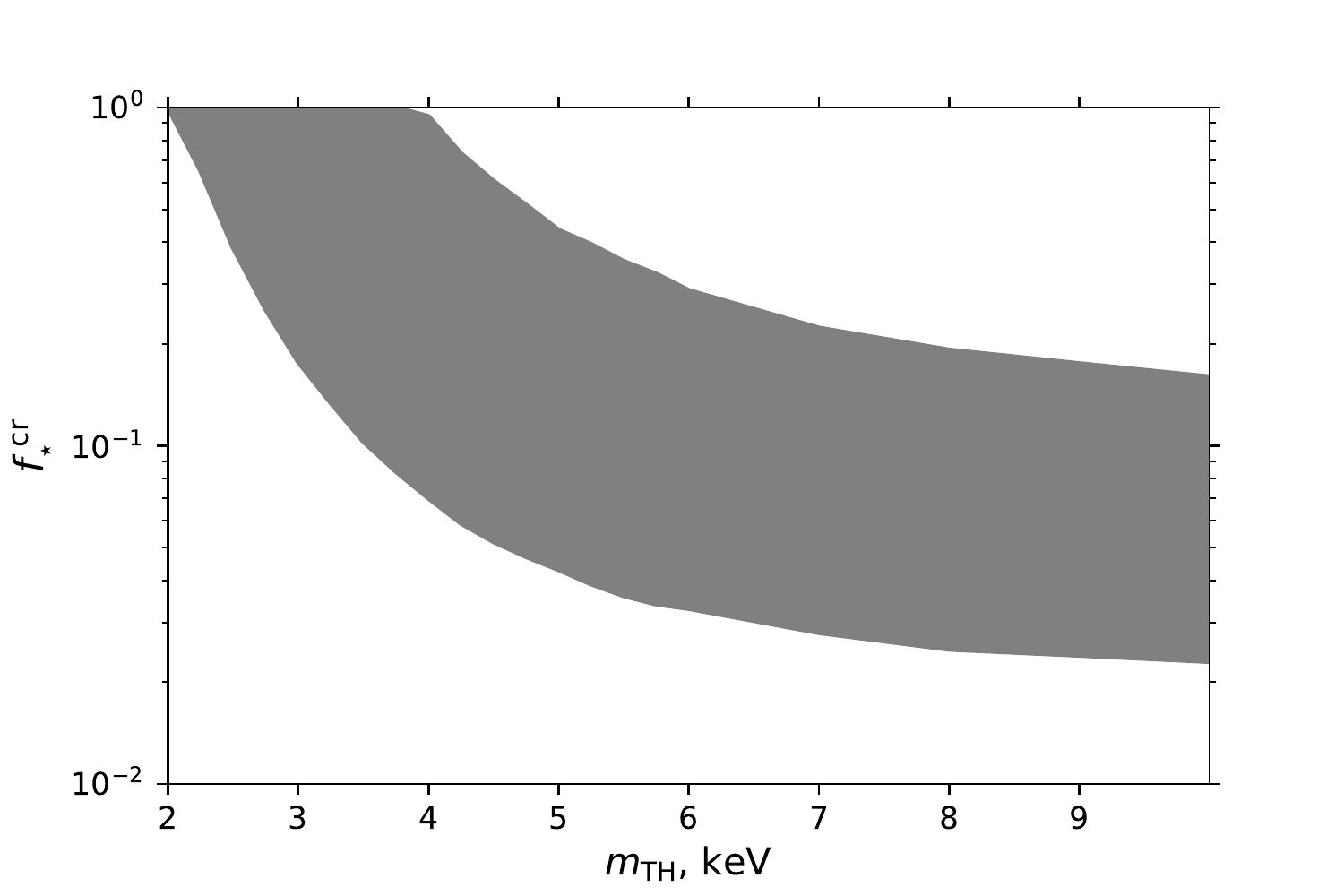}
    \caption{The range of values of $f_*$ for which the minimum of the
      absorption trough lies in the redshift range $15.8 \le z \le 18.7$,
      consistent with EDGES observations. For all models the minimal virial temperature of halos is fixed at $T_{vir} = 10^4$~K, corresponding to atomic hydrogen cooling}
    \label{fig:f_star_region}
\end{figure}

In Fig.~\ref{fig:f_star_region} we plot the range of $f_*$'s that have the minimum of the absorption trough for $15.8 \le z \le 18.7$. We see that starting from $m_{\text{TR}} \le 4$~keV $f_*$ can be as large as 100\% and that for masses of this order or above.
Given several orders of magnitude uncertainties in $f_*$ (as discussed above), the only robust bound can be obtained if one chooses $f_* = 1$; at most all baryons enter star formation.

\begin{figure*}
    \centering
    \includegraphics[width=0.6\textwidth]{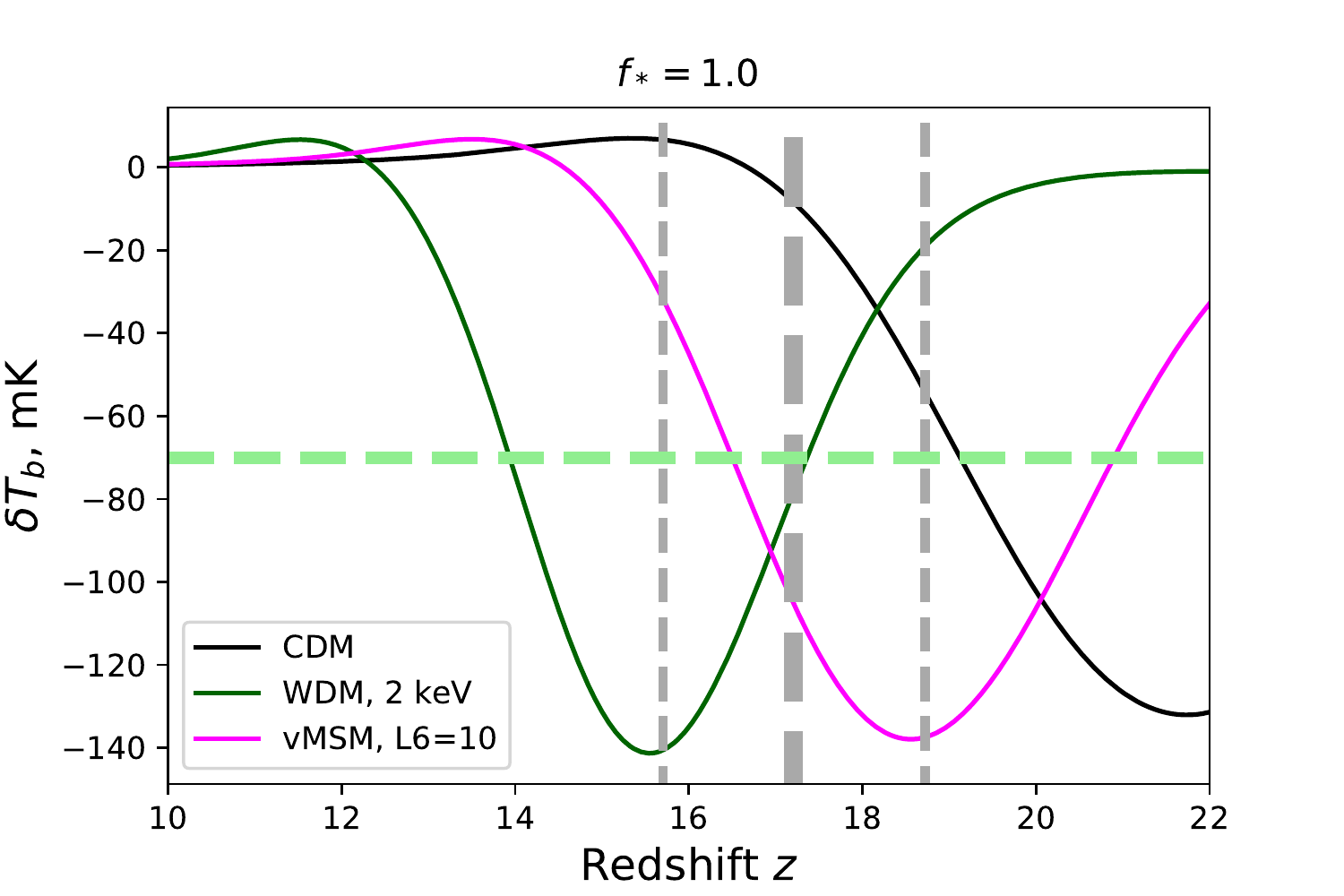}
    \caption{The same as in Figure~\protect\ref{fig:dTb} but for $f_* = 1.0$. As we see, even the thermal WDM model with particle mass $m_\text{TH} = 2$~keV is consistent with observations.}
    \label{fig:dTb-1}
\end{figure*}
In this case, for example, thermal WDM masses as light as $m_\text{TR} \ge
2$~keV cannot be excluded (see Fig.~\ref{fig:dTb-1}).
This puts the sensitivity of the EDGES signal in line with a number of previous bounds on WDM parameters (see, e.g., the Lyman-$\alpha$ constraints~\cite{Garzilli:2015iwa}, taking into account proper marginalizations over possible thermal histories;  bounds \cite{Menci:2016eui} from counting of high-$z$ galaxies; bounds \cite{Birrer:2017rpp,Vegetti:2018dly} from strong gravitational lensing; bounds~\cite{Lovell:2013ola,Kennedy:2013uta} from the Milky Way satellite counts, etc.).
As \cite{Mesinger:2012ys} demonstrates, future measurements of star-formation efficiency at high redshifts, as well as the 21-cm power spectrum, are required to improve the sensitivity for WDM particles.

\bigskip

In this paper we have concentrated  on the redshift position of minimum of $\delta T_b(z)$ as an indicator of star-forming processes at  high-redshifts.
However, both the depth of the 21-cm absorption trough and its width carry important information about the underlying physics.

\begin{figure*}
    \centering
    \includegraphics[width=0.48\textwidth]{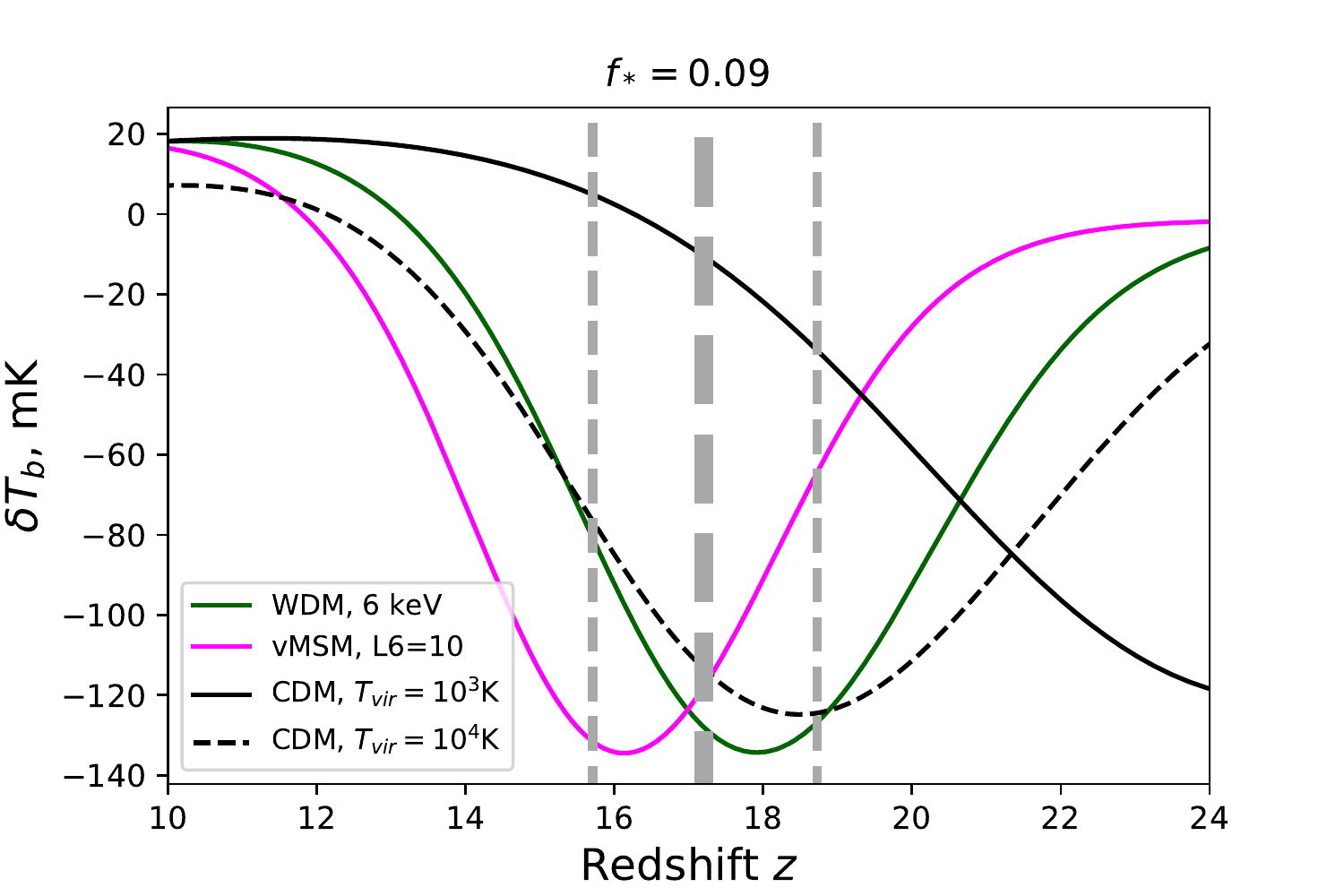}
    \includegraphics[width=0.48\textwidth]{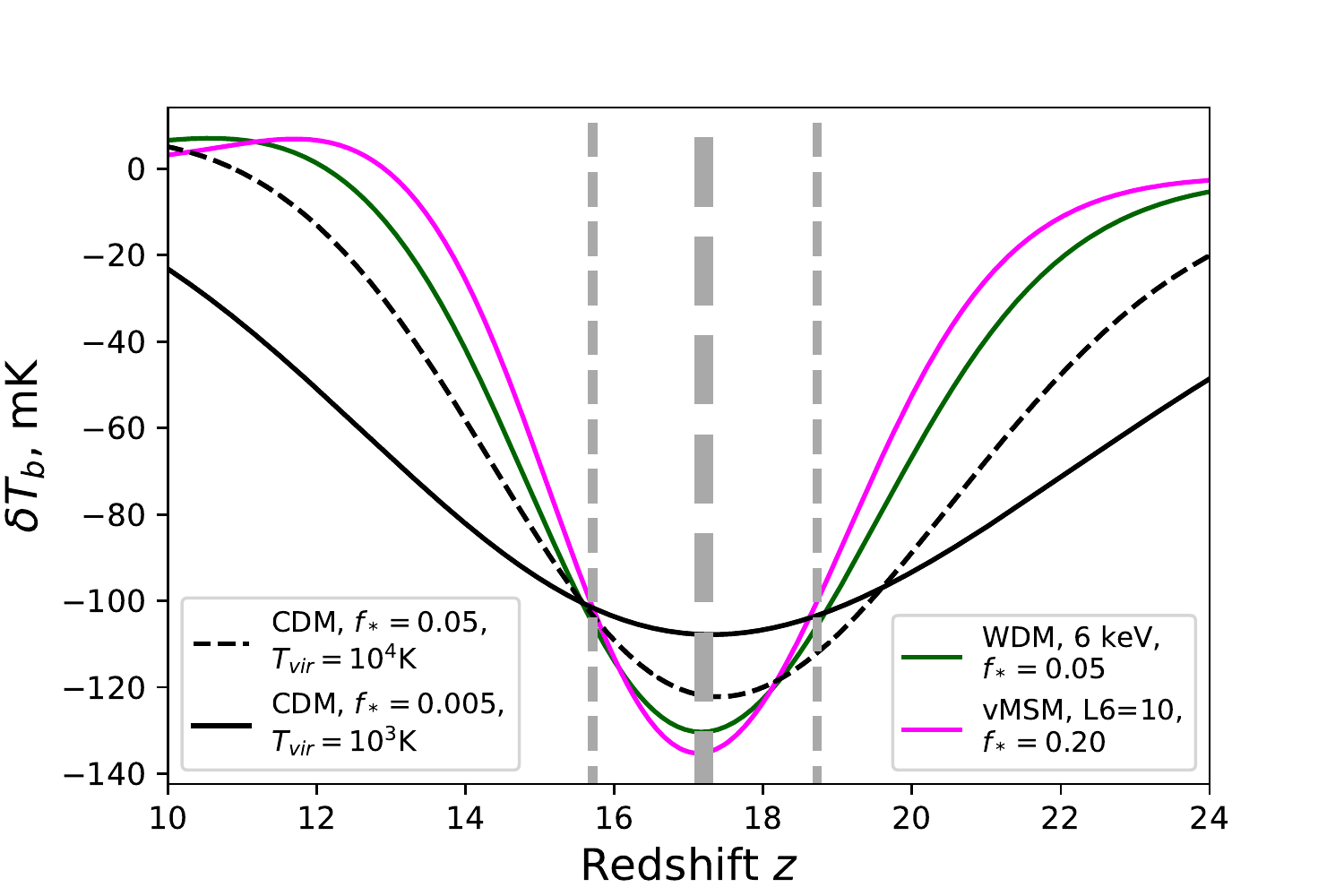}
    \caption{\textit{Left panel}: The same as in Fig.~\protect\ref{fig:dTb} but with minimal halo temperature $T_{vir} = 10^3$~K, which corresponds to molecular hydrogen cooling; see, e.g., \cite{Galli:1998dh} and Fig.~12 of~\cite{Barkana:2000fd} for details.
      By comparing with Fig.~\protect\ref{fig:dTb} we see that the decrease of $T_{vir}$ from $10^4$~K to $10^3$~K essentially does not change absorption profiles for 6-keV WDM and $\text{L}6=10$ $\nu$MSM models.
      In contrast, in the CDM model, predictions change dramatically both in profile width and position of the minimum (here, the black solid curve denotes $T_{vir} = 10^3$~K while black dashed curve denotes $T_{vir} = 10^4$~K), owing to a significant number of small mass halos (cf.~\protect\cite{Leo:2019gwh}).
      \textit{Right panel}: The same profiles as in the left panel but with values of $f_*$ adjusted to match the best-fit position of the EDGES absorption trough.}
    \label{fig:dTb-009-T1000}
\end{figure*}

Much like the position, the width of the obtained profile also depends on the cosmology.  
When using $T_{vir} = 10^3$~K (molecular cooling) and ignoring possible suppression due to the Lyman-Werner radiation background (see, e.g.,~\cite{Yue:2012na}), we see that CDM predicts an absorption-trough width which is larger than the one observed by the EDGES experiment, Fig.~\ref{fig:dTb-009-T1000}. For the WDM and $\nu$MSM profiles the molecular cooling brings little to no effect due to the lack of substructures of the mass $\sim M_{\rm min}$.

The depth of the observed trough is much greater than what any of the models discussed in this paper predict. To date, only additional nongravitational baryon-DM interactions can accommodate such a strong spin-temperature cooling, which is beyond the scope of this paper~\cite{Barkana:2018lgd,Berlin:2018sjs,Munoz:2018pzp,Fialkov:2018xre}.

\bigskip

\textit{To summarize}, we discussed the large uncertainty in star formation at very high redshifts ($z\sim 17$), which are probed by recent EDGES observations of the global 21-cm signal. As a consequence, using only this signal it is impossible to robustly constrain the parameters of dark matter models, such as the mass of the warm dark matter particle. Conversely, various DM models need distinct star-formation scenarios to fit the signal. Detailed future studies of star formation at very high redshifts ($z \gtrsim 10$), together with detailed modeling of structure assembly and early star formation, will reduce the existing uncertainties. Ongoing and future studies of the 21-cm signal remain promising tools for inferring the key dark matter parameters.

\bigskip

\textit{Acknowledgements.} We thank Tom Theuns for valuable comments on an earlier version of this paper and the authors of~\cite{Leo:2019gwh} for sharing with us results of their work before publication.
The work of D.I. and O.R. œwas supported by the Carlsberg Foundation. The work of A.R.\ was partially supported by grant for Young Scientists Research Laboratories
of the National Academy of Sciences of Ukraine. A.R.\ also acknowledges the grant from the Abdus Salam International Centre for Theoretical Physics, Trieste, Italy. 
This project has received funding from the European Research Council (ERC) under the European Union's Horizon 2020 Research and Innovation Programme (GA 694896).

\bibliography{preamble,refs}

\let\jnlstyle=\rm\def\jref#1{{\jnlstyle#1}}\def\aj{\jref{AJ}}
  \def\araa{\jref{ARA\&A}} \def\apj{\jref{ApJ}\ } \def\apjl{\jref{ApJ}\ }
  \def\apjs{\jref{ApJS}} \def\ao{\jref{Appl.~Opt.}} \def\apss{\jref{Ap\&SS}}
  \def\aap{\jref{A\&A}} \def\aapr{\jref{A\&A~Rev.}} \def\aaps{\jref{A\&AS}}
  \def\azh{\jref{AZh}} \def\baas{\jref{BAAS}} \def\jrasc{\jref{JRASC}}
  \def\memras{\jref{MmRAS}} \def\mnras{\jref{MNRAS}\ }
  \def\pra{\jref{Phys.~Rev.~A}\ } \def\prb{\jref{Phys.~Rev.~B}\ }
  \def\prc{\jref{Phys.~Rev.~C}\ } \def\prd{\jref{Phys.~Rev.~D}\ }
  \def\pre{\jref{Phys.~Rev.~E}} \def\prl{\jref{Phys.~Rev.~Lett.}}
  \def\pasp{\jref{PASP}} \def\pasj{\jref{PASJ}} \def\qjras{\jref{QJRAS}}
  \def\skytel{\jref{S\&T}} \def\solphys{\jref{Sol.~Phys.}}
  \def\sovast{\jref{Soviet~Ast.}} \def\ssr{\jref{Space~Sci.~Rev.}}
  \def\zap{\jref{ZAp}} \def\nat{\jref{Nature}\ } \def\iaucirc{\jref{IAU~Circ.}}
  \def\aplett{\jref{Astrophys.~Lett.}}
  \def\apspr{\jref{Astrophys.~Space~Phys.~Res.}}
  \def\bain{\jref{Bull.~Astron.~Inst.~Netherlands}}
  \def\fcp{\jref{Fund.~Cosmic~Phys.}} \def\gca{\jref{Geochim.~Cosmochim.~Acta}}
  \def\grl{\jref{Geophys.~Res.~Lett.}} \def\jcp{\jref{J.~Chem.~Phys.}}
  \def\jgr{\jref{J.~Geophys.~Res.}}
  \def\jqsrt{\jref{J.~Quant.~Spec.~Radiat.~Transf.}}
  \def\memsai{\jref{Mem.~Soc.~Astron.~Italiana}}
  \def\nphysa{\jref{Nucl.~Phys.~A}} \def\physrep{\jref{Phys.~Rep.}}
  \def\physscr{\jref{Phys.~Scr}} \def\planss{\jref{Planet.~Space~Sci.}}
  \def\procspie{\jref{Proc.~SPIE}} \let\astap=\aap \let\apjlett=\apjl
  \let\apjsupp=\apjs \let\applopt=\ao \def\jcap{\jref{JCAP}}
\begin{thebibliography}{106}%
\makeatletter
\providecommand \@ifxundefined [1]{%
 \@ifx{#1\undefined}
}%
\providecommand \@ifnum [1]{%
 \ifnum #1\expandafter \@firstoftwo
 \else \expandafter \@secondoftwo
 \fi
}%
\providecommand \@ifx [1]{%
 \ifx #1\expandafter \@firstoftwo
 \else \expandafter \@secondoftwo
 \fi
}%
\providecommand \natexlab [1]{#1}%
\providecommand \enquote  [1]{``#1''}%
\providecommand \bibnamefont  [1]{#1}%
\providecommand \bibfnamefont [1]{#1}%
\providecommand \citenamefont [1]{#1}%
\providecommand \href@noop [0]{\@secondoftwo}%
\providecommand \href [0]{\begingroup \@sanitize@url \@href}%
\providecommand \@href[1]{\@@startlink{#1}\@@href}%
\providecommand \@@href[1]{\endgroup#1\@@endlink}%
\providecommand \@sanitize@url [0]{\catcode `\\12\catcode `\$12\catcode
  `\&12\catcode `\#12\catcode `\^12\catcode `\_12\catcode `\%12\relax}%
\providecommand \@@startlink[1]{}%
\providecommand \@@endlink[0]{}%
\providecommand \url  [0]{\begingroup\@sanitize@url \@url }%
\providecommand \@url [1]{\endgroup\@href {#1}{\urlprefix }}%
\providecommand \urlprefix  [0]{URL }%
\providecommand \Eprint [0]{\href }%
\providecommand \doibase [0]{http://dx.doi.org/}%
\providecommand \selectlanguage [0]{\@gobble}%
\providecommand \bibinfo  [0]{\@secondoftwo}%
\providecommand \bibfield  [0]{\@secondoftwo}%
\providecommand \translation [1]{[#1]}%
\providecommand \BibitemOpen [0]{}%
\providecommand \bibitemStop [0]{}%
\providecommand \bibitemNoStop [0]{.\EOS\space}%
\providecommand \EOS [0]{\spacefactor3000\relax}%
\providecommand \BibitemShut  [1]{\csname bibitem#1\endcsname}%
\let\auto@bib@innerbib\@empty
\bibitem [{\citenamefont {Patil}\ \emph {et~al.}(2017)\citenamefont {Patil}
  \emph {et~al.}}]{Patil:2017zqk}%
  \BibitemOpen
  \bibfield  {author} {\bibinfo {author} {\bibfnamefont {A.~H.}\ \bibnamefont
  {Patil}} \emph {et~al.},\ }\bibfield  {title} {\enquote {\bibinfo {title}
  {{Upper limits on the 21-cm Epoch of Reionization power spectrum from one
  night with LOFAR}},}\ }\href {\doibase 10.3847/1538-4357/aa63e7} {\bibfield
  {journal} {\bibinfo  {journal} {Astrophys. J.}\ }\textbf {\bibinfo {volume}
  {838}},\ \bibinfo {pages} {65} (\bibinfo {year} {2017})},\ \Eprint
  {http://arxiv.org/abs/1702.08679} {arXiv:1702.08679 [astro-ph.CO]}
  \BibitemShut {NoStop}%
\bibitem [{\citenamefont {{Gehlot}}\ \emph {et~al.}(2019)\citenamefont
  {{Gehlot}}, \citenamefont {{Mertens}}, \citenamefont {{Koopmans}},
  \citenamefont {{Brentjens}}, \citenamefont {{Zaroubi}}, \citenamefont
  {{Ciardi}}, \citenamefont {{Ghosh}}, \citenamefont {{Hatef}}, \citenamefont
  {{Iliev}}, \citenamefont {{Jeli{\'c}}}, \citenamefont {{}}, \citenamefont
  {{Kooistra}}, \citenamefont {{Krause}}, \citenamefont {{Mellema}},
  \citenamefont {{Mevius}}, \citenamefont {{Mitra}}, \citenamefont
  {{Offringa}}, \citenamefont {{Pandey}}, \citenamefont {{Sardarabadi}},
  \citenamefont {{Schaye}}, \citenamefont {{Silva}}, \citenamefont
  {{Vedantham}},\ and\ \citenamefont {{Yatawatta}}}]{Gehlot:18}%
  \BibitemOpen
  \bibfield  {author} {\bibinfo {author} {\bibfnamefont {B.~K.}\ \bibnamefont
  {{Gehlot}}}, \bibinfo {author} {\bibfnamefont {F.~G.}\ \bibnamefont
  {{Mertens}}}, \bibinfo {author} {\bibfnamefont {L.~V.~E.}\ \bibnamefont
  {{Koopmans}}}, \bibinfo {author} {\bibfnamefont {M.~A.}\ \bibnamefont
  {{Brentjens}}}, \bibinfo {author} {\bibfnamefont {S.}~\bibnamefont
  {{Zaroubi}}}, \bibinfo {author} {\bibfnamefont {B.}~\bibnamefont {{Ciardi}}},
  \bibinfo {author} {\bibfnamefont {A.}~\bibnamefont {{Ghosh}}}, \bibinfo
  {author} {\bibfnamefont {M.}~\bibnamefont {{Hatef}}}, \bibinfo {author}
  {\bibfnamefont {I.~T.}\ \bibnamefont {{Iliev}}}, \bibinfo {author}
  {\bibnamefont {{Jeli{\'c}}}}, \bibinfo {author} {\bibfnamefont
  {V.}~\bibnamefont {{}}}, \bibinfo {author} {\bibfnamefont {R.}~\bibnamefont
  {{Kooistra}}}, \bibinfo {author} {\bibfnamefont {F.}~\bibnamefont
  {{Krause}}}, \bibinfo {author} {\bibfnamefont {G.}~\bibnamefont {{Mellema}}},
  \bibinfo {author} {\bibfnamefont {M.}~\bibnamefont {{Mevius}}}, \bibinfo
  {author} {\bibfnamefont {M.}~\bibnamefont {{Mitra}}}, \bibinfo {author}
  {\bibfnamefont {A.~R.}\ \bibnamefont {{Offringa}}}, \bibinfo {author}
  {\bibfnamefont {V.~N.}\ \bibnamefont {{Pandey}}}, \bibinfo {author}
  {\bibfnamefont {A.~M.}\ \bibnamefont {{Sardarabadi}}}, \bibinfo {author}
  {\bibfnamefont {J.}~\bibnamefont {{Schaye}}}, \bibinfo {author}
  {\bibfnamefont {M.~B.}\ \bibnamefont {{Silva}}}, \bibinfo {author}
  {\bibfnamefont {H.~K.}\ \bibnamefont {{Vedantham}}}, \ and\ \bibinfo {author}
  {\bibfnamefont {S.}~\bibnamefont {{Yatawatta}}},\ }\bibfield  {title}
  {\enquote {\bibinfo {title} {{The first power spectrum limit on the 21-cm
  signal of neutral hydrogen during the Cosmic Dawn at z = 20-25 from
  LOFAR}},}\ }\href {\doibase 10.1093/mnras/stz1937} {\bibfield  {journal}
  {\bibinfo  {journal} {\mnras}\ }\textbf {\bibinfo {volume} {488}},\ \bibinfo
  {pages} {4271--4287} (\bibinfo {year} {2019})},\ \Eprint
  {http://arxiv.org/abs/1809.06661} {arXiv:1809.06661 [astro-ph.IM]}
  \BibitemShut {NoStop}%
\bibitem [{\citenamefont {Paciga}\ \emph {et~al.}(2013)\citenamefont {Paciga}
  \emph {et~al.}}]{Paciga:2013fj}%
  \BibitemOpen
  \bibfield  {author} {\bibinfo {author} {\bibfnamefont {Gregory}\ \bibnamefont
  {Paciga}} \emph {et~al.},\ }\bibfield  {title} {\enquote {\bibinfo {title}
  {{A refined foreground-corrected limit on the HI power spectrum at z=8.6 from
  the GMRT Epoch of Reionization Experiment}},}\ }\href {\doibase
  10.1093/mnras/stt753} {\bibfield  {journal} {\bibinfo  {journal} {Mon. Not.
  Roy. Astron. Soc.}\ }\textbf {\bibinfo {volume} {433}},\ \bibinfo {pages}
  {639} (\bibinfo {year} {2013})},\ \Eprint {http://arxiv.org/abs/1301.5906}
  {arXiv:1301.5906 [astro-ph.CO]} \BibitemShut {NoStop}%
\bibitem [{\citenamefont {Ali}\ \emph {et~al.}(2015)\citenamefont {Ali} \emph
  {et~al.}}]{Ali:2015uua}%
  \BibitemOpen
  \bibfield  {author} {\bibinfo {author} {\bibfnamefont {Zaki~S.}\ \bibnamefont
  {Ali}} \emph {et~al.},\ }\bibfield  {title} {\enquote {\bibinfo {title}
  {{PAPER-64 Constraints on Reionization: The 21cm Power Spectrum at
  $z=$8.4}},}\ }\href {\doibase 10.1088/0004-637X/809/1/61} {\bibfield
  {journal} {\bibinfo  {journal} {Astrophys. J.}\ }\textbf {\bibinfo {volume}
  {809}},\ \bibinfo {pages} {61} (\bibinfo {year} {2015})},\ \Eprint
  {http://arxiv.org/abs/1502.06016} {arXiv:1502.06016 [astro-ph.CO]}
  \BibitemShut {NoStop}%
\bibitem [{\citenamefont {{Ali}}\ \emph {et~al.}(2018)\citenamefont {{Ali}},
  \citenamefont {{Parsons}}, \citenamefont {{Zheng}}, \citenamefont {{Pober}},
  \citenamefont {{Liu}}, \citenamefont {{Aguirre}}, \citenamefont {{Bradley}},
  \citenamefont {{Bernardi}}, \citenamefont {{Carilli}}, \citenamefont
  {{Cheng}}, \citenamefont {{DeBoer}}, \citenamefont {{Dexter}}, \citenamefont
  {{Grobbelaar}}, \citenamefont {{Horrell}}, \citenamefont {{Jacobs}},
  \citenamefont {{Klima}}, \citenamefont {{MacMahon}}, \citenamefont {{Maree}},
  \citenamefont {{Moore}}, \citenamefont {{Razavi}}, \citenamefont {{Stefan}},
  \citenamefont {{Walbrugh}},\ and\ \citenamefont {{Walker}}}]{Ali:18}%
  \BibitemOpen
  \bibfield  {author} {\bibinfo {author} {\bibfnamefont {Z.~S.}\ \bibnamefont
  {{Ali}}}, \bibinfo {author} {\bibfnamefont {A.~R.}\ \bibnamefont
  {{Parsons}}}, \bibinfo {author} {\bibfnamefont {H.}~\bibnamefont {{Zheng}}},
  \bibinfo {author} {\bibfnamefont {J.~C.}\ \bibnamefont {{Pober}}}, \bibinfo
  {author} {\bibfnamefont {A.}~\bibnamefont {{Liu}}}, \bibinfo {author}
  {\bibfnamefont {J.~E.}\ \bibnamefont {{Aguirre}}}, \bibinfo {author}
  {\bibfnamefont {R.~F.}\ \bibnamefont {{Bradley}}}, \bibinfo {author}
  {\bibfnamefont {G.}~\bibnamefont {{Bernardi}}}, \bibinfo {author}
  {\bibfnamefont {C.~L.}\ \bibnamefont {{Carilli}}}, \bibinfo {author}
  {\bibfnamefont {C.}~\bibnamefont {{Cheng}}}, \bibinfo {author} {\bibfnamefont
  {D.~R.}\ \bibnamefont {{DeBoer}}}, \bibinfo {author} {\bibfnamefont {M.~R.}\
  \bibnamefont {{Dexter}}}, \bibinfo {author} {\bibfnamefont {J.}~\bibnamefont
  {{Grobbelaar}}}, \bibinfo {author} {\bibfnamefont {J.}~\bibnamefont
  {{Horrell}}}, \bibinfo {author} {\bibfnamefont {D.~C.}\ \bibnamefont
  {{Jacobs}}}, \bibinfo {author} {\bibfnamefont {P.}~\bibnamefont {{Klima}}},
  \bibinfo {author} {\bibfnamefont {D.~H.~E.}\ \bibnamefont {{MacMahon}}},
  \bibinfo {author} {\bibfnamefont {M.}~\bibnamefont {{Maree}}}, \bibinfo
  {author} {\bibfnamefont {D.~F.}\ \bibnamefont {{Moore}}}, \bibinfo {author}
  {\bibfnamefont {N.}~\bibnamefont {{Razavi}}}, \bibinfo {author}
  {\bibfnamefont {I.~I.}\ \bibnamefont {{Stefan}}}, \bibinfo {author}
  {\bibfnamefont {W.~P.}\ \bibnamefont {{Walbrugh}}}, \ and\ \bibinfo {author}
  {\bibfnamefont {A.}~\bibnamefont {{Walker}}},\ }\bibfield  {title} {\enquote
  {\bibinfo {title} {{Erratum: PAPER64 Constraints on Reionization: The 21 cm
  Power Spectrum at z = 8.4}},}\ }\href {\doibase 10.3847/1538-4357/aad7b4}
  {\bibfield  {journal} {\bibinfo  {journal} {\apj}\ }\textbf {\bibinfo
  {volume} {863}},\ \bibinfo {eid} {201} (\bibinfo {year} {2018})}\BibitemShut
  {NoStop}%
\bibitem [{\citenamefont {Ewall-Wice}\ \emph {et~al.}(2017)\citenamefont
  {Ewall-Wice}, \citenamefont {Dillon}, \citenamefont {Liu},\ and\
  \citenamefont {Hewitt}}]{Ewall-Wice:2016bhu}%
  \BibitemOpen
  \bibfield  {author} {\bibinfo {author} {\bibfnamefont {Aaron}\ \bibnamefont
  {Ewall-Wice}}, \bibinfo {author} {\bibfnamefont {Joshua~S.}\ \bibnamefont
  {Dillon}}, \bibinfo {author} {\bibfnamefont {Adrian}\ \bibnamefont {Liu}}, \
  and\ \bibinfo {author} {\bibfnamefont {Jacqueline}\ \bibnamefont {Hewitt}},\
  }\bibfield  {title} {\enquote {\bibinfo {title} {{The impact of modelling
  errors on interferometer calibration for 21 cm power spectra}},}\ }\href
  {\doibase 10.1093/mnras/stx1221} {\bibfield  {journal} {\bibinfo  {journal}
  {Mon. Not. Roy. Astron. Soc.}\ }\textbf {\bibinfo {volume} {470}},\ \bibinfo
  {pages} {1849--1870} (\bibinfo {year} {2017})},\ \Eprint
  {http://arxiv.org/abs/1610.02689} {arXiv:1610.02689 [astro-ph.CO]}
  \BibitemShut {NoStop}%
\bibitem [{\citenamefont {Bowman}\ \emph {et~al.}(2018)\citenamefont {Bowman},
  \citenamefont {Rogers}, \citenamefont {Monsalve}, \citenamefont {Mozdzen},\
  and\ \citenamefont {Mahesh}}]{Bowman:2018yin}%
  \BibitemOpen
  \bibfield  {author} {\bibinfo {author} {\bibfnamefont {Judd~D.}\ \bibnamefont
  {Bowman}}, \bibinfo {author} {\bibfnamefont {Alan E.~E.}\ \bibnamefont
  {Rogers}}, \bibinfo {author} {\bibfnamefont {Raul~A.}\ \bibnamefont
  {Monsalve}}, \bibinfo {author} {\bibfnamefont {Thomas~J.}\ \bibnamefont
  {Mozdzen}}, \ and\ \bibinfo {author} {\bibfnamefont {Nivedita}\ \bibnamefont
  {Mahesh}},\ }\bibfield  {title} {\enquote {\bibinfo {title} {{An absorption
  profile centred at 78 megahertz in the sky-averaged spectrum}},}\ }\href
  {\doibase 10.1038/nature25792} {\bibfield  {journal} {\bibinfo  {journal}
  {Nature}\ }\textbf {\bibinfo {volume} {555}},\ \bibinfo {pages} {67--70}
  (\bibinfo {year} {2018})}\BibitemShut {NoStop}%
\bibitem [{\citenamefont {Hills}\ \emph {et~al.}(2018)\citenamefont {Hills},
  \citenamefont {Kulkarni}, \citenamefont {Meerburg},\ and\ \citenamefont
  {Puchwein}}]{Hills:2018vyr}%
  \BibitemOpen
  \bibfield  {author} {\bibinfo {author} {\bibfnamefont {Richard}\ \bibnamefont
  {Hills}}, \bibinfo {author} {\bibfnamefont {Girish}\ \bibnamefont
  {Kulkarni}}, \bibinfo {author} {\bibfnamefont {P.~Daniel}\ \bibnamefont
  {Meerburg}}, \ and\ \bibinfo {author} {\bibfnamefont {Ewald}\ \bibnamefont
  {Puchwein}},\ }\bibfield  {title} {\enquote {\bibinfo {title} {{Concerns
  about modelling of the EDGES data}},}\ }\href {\doibase
  10.1038/s41586-018-0796-5} {\bibfield  {journal} {\bibinfo  {journal}
  {Nature}\ }\textbf {\bibinfo {volume} {564}},\ \bibinfo {pages} {E32--E34}
  (\bibinfo {year} {2018})},\ \Eprint {http://arxiv.org/abs/1805.01421}
  {arXiv:1805.01421 [astro-ph.CO]} \BibitemShut {NoStop}%
\bibitem [{\citenamefont {Bradley}\ \emph {et~al.}(2019)\citenamefont
  {Bradley}, \citenamefont {Tauscher}, \citenamefont {Rapetti},\ and\
  \citenamefont {Burns}}]{Bradley:2018eev}%
  \BibitemOpen
  \bibfield  {author} {\bibinfo {author} {\bibfnamefont {Richard~F.}\
  \bibnamefont {Bradley}}, \bibinfo {author} {\bibfnamefont {Keith}\
  \bibnamefont {Tauscher}}, \bibinfo {author} {\bibfnamefont {David}\
  \bibnamefont {Rapetti}}, \ and\ \bibinfo {author} {\bibfnamefont {Jack~O.}\
  \bibnamefont {Burns}},\ }\bibfield  {title} {\enquote {\bibinfo {title} {{A
  Ground Plane Artifact that Induces an Absorption Profile in Averaged Spectra
  from Global 21-cm Measurements - with Possible Application to EDGES}},}\
  }\href {\doibase 10.3847/1538-4357/ab0d8b} {\bibfield  {journal} {\bibinfo
  {journal} {Astrophys. J.}\ }\textbf {\bibinfo {volume} {874}},\ \bibinfo
  {pages} {153} (\bibinfo {year} {2019})},\ \Eprint
  {http://arxiv.org/abs/1810.09015} {arXiv:1810.09015 [astro-ph.IM]}
  \BibitemShut {NoStop}%
\bibitem [{\citenamefont {DeBoer}\ \emph {et~al.}(2017)\citenamefont {DeBoer}
  \emph {et~al.}}]{DeBoer:2016tnn}%
  \BibitemOpen
  \bibfield  {author} {\bibinfo {author} {\bibfnamefont {David~R.}\
  \bibnamefont {DeBoer}} \emph {et~al.},\ }\bibfield  {title} {\enquote
  {\bibinfo {title} {{Hydrogen Epoch of Reionization Array (HERA)}},}\ }\href
  {\doibase 10.1088/1538-3873/129/974/045001} {\bibfield  {journal} {\bibinfo
  {journal} {Publ. Astron. Soc. Pac.}\ }\textbf {\bibinfo {volume} {129}},\
  \bibinfo {pages} {045001} (\bibinfo {year} {2017})},\ \Eprint
  {http://arxiv.org/abs/1606.07473} {arXiv:1606.07473 [astro-ph.IM]}
  \BibitemShut {NoStop}%
\bibitem [{\citenamefont {Koopmans}\ \emph {et~al.}(2015)\citenamefont
  {Koopmans} \emph {et~al.}}]{Koopmans:2015sua}%
  \BibitemOpen
  \bibfield  {author} {\bibinfo {author} {\bibfnamefont {L.~V.~E.}\
  \bibnamefont {Koopmans}} \emph {et~al.},\ }\bibfield  {title} {\enquote
  {\bibinfo {title} {{The Cosmic Dawn and Epoch of Reionization with the Square
  Kilometre Array}},}\ }\bibfield  {booktitle} {\emph {\bibinfo {booktitle}
  {{Proceedings, Advancing Astrophysics with the Square Kilometre Array
  (AASKA14): Giardini Naxos, Italy, June 9-13, 2014}}},\ }\href {\doibase
  10.22323/1.215.0001} {\bibfield  {journal} {\bibinfo  {journal} {PoS}\
  }\textbf {\bibinfo {volume} {AASKA14}},\ \bibinfo {pages} {001} (\bibinfo
  {year} {2015})},\ \Eprint {http://arxiv.org/abs/1505.07568} {arXiv:1505.07568
  [astro-ph.CO]} \BibitemShut {NoStop}%
\bibitem [{\citenamefont {Bull}\ \emph {et~al.}(2018)\citenamefont {Bull} \emph
  {et~al.}}]{Bull:2018lat}%
  \BibitemOpen
  \bibfield  {author} {\bibinfo {author} {\bibfnamefont {P.}~\bibnamefont
  {Bull}} \emph {et~al.},\ }\bibfield  {title} {\enquote {\bibinfo {title}
  {{Fundamental Physics with the Square Kilometer Array}},}\ }\href@noop {} {\
  (\bibinfo {year} {2018})},\ \Eprint {http://arxiv.org/abs/1810.02680}
  {arXiv:1810.02680 [astro-ph.CO]} \BibitemShut {NoStop}%
\bibitem [{\citenamefont {McQuinn}\ \emph {et~al.}(2006)\citenamefont
  {McQuinn}, \citenamefont {Zahn}, \citenamefont {Zaldarriaga}, \citenamefont
  {Hernquist},\ and\ \citenamefont {Furlanetto}}]{McQuinn:2005hk}%
  \BibitemOpen
  \bibfield  {author} {\bibinfo {author} {\bibfnamefont {Matthew}\ \bibnamefont
  {McQuinn}}, \bibinfo {author} {\bibfnamefont {Oliver}\ \bibnamefont {Zahn}},
  \bibinfo {author} {\bibfnamefont {Matias}\ \bibnamefont {Zaldarriaga}},
  \bibinfo {author} {\bibfnamefont {Lars}\ \bibnamefont {Hernquist}}, \ and\
  \bibinfo {author} {\bibfnamefont {Steven~R.}\ \bibnamefont {Furlanetto}},\
  }\bibfield  {title} {\enquote {\bibinfo {title} {{Cosmological parameter
  estimation using 21 cm radiation from the epoch of reionization}},}\ }\href
  {\doibase 10.1086/505167} {\bibfield  {journal} {\bibinfo  {journal}
  {Astrophys. J.}\ }\textbf {\bibinfo {volume} {653}},\ \bibinfo {pages}
  {815--830} (\bibinfo {year} {2006})},\ \Eprint
  {http://arxiv.org/abs/astro-ph/0512263} {arXiv:astro-ph/0512263 [astro-ph]}
  \BibitemShut {NoStop}%
\bibitem [{\citenamefont {Mao}\ \emph {et~al.}(2008)\citenamefont {Mao},
  \citenamefont {Tegmark}, \citenamefont {McQuinn}, \citenamefont
  {Zaldarriaga},\ and\ \citenamefont {Zahn}}]{Mao:2008ug}%
  \BibitemOpen
  \bibfield  {author} {\bibinfo {author} {\bibfnamefont {Yi}~\bibnamefont
  {Mao}}, \bibinfo {author} {\bibfnamefont {Max}\ \bibnamefont {Tegmark}},
  \bibinfo {author} {\bibfnamefont {Matthew}\ \bibnamefont {McQuinn}}, \bibinfo
  {author} {\bibfnamefont {Matias}\ \bibnamefont {Zaldarriaga}}, \ and\
  \bibinfo {author} {\bibfnamefont {Oliver}\ \bibnamefont {Zahn}},\ }\bibfield
  {title} {\enquote {\bibinfo {title} {{How accurately can 21 cm tomography
  constrain cosmology?}}}\ }\href {\doibase 10.1103/PhysRevD.78.023529}
  {\bibfield  {journal} {\bibinfo  {journal} {Phys. Rev.}\ }\textbf {\bibinfo
  {volume} {D78}},\ \bibinfo {pages} {023529} (\bibinfo {year} {2008})},\
  \Eprint {http://arxiv.org/abs/0802.1710} {arXiv:0802.1710 [astro-ph]}
  \BibitemShut {NoStop}%
\bibitem [{\citenamefont {Oyama}\ \emph {et~al.}(2016)\citenamefont {Oyama},
  \citenamefont {Kohri},\ and\ \citenamefont {Hazumi}}]{Oyama:2015gma}%
  \BibitemOpen
  \bibfield  {author} {\bibinfo {author} {\bibfnamefont {Yoshihiko}\
  \bibnamefont {Oyama}}, \bibinfo {author} {\bibfnamefont {Kazunori}\
  \bibnamefont {Kohri}}, \ and\ \bibinfo {author} {\bibfnamefont {Masashi}\
  \bibnamefont {Hazumi}},\ }\bibfield  {title} {\enquote {\bibinfo {title}
  {{Constraints on the neutrino parameters by future cosmological 21 cm line
  and precise CMB polarization observations}},}\ }\href {\doibase
  10.1088/1475-7516/2016/02/008} {\bibfield  {journal} {\bibinfo  {journal}
  {JCAP}\ }\textbf {\bibinfo {volume} {1602}},\ \bibinfo {pages} {008}
  (\bibinfo {year} {2016})},\ \Eprint {http://arxiv.org/abs/1510.03806}
  {arXiv:1510.03806 [astro-ph.CO]} \BibitemShut {NoStop}%
\bibitem [{\citenamefont {D'Amico}\ \emph {et~al.}(2018)\citenamefont
  {D'Amico}, \citenamefont {Panci},\ and\ \citenamefont
  {Strumia}}]{DAmico:2018sxd}%
  \BibitemOpen
  \bibfield  {author} {\bibinfo {author} {\bibfnamefont {Guido}\ \bibnamefont
  {D'Amico}}, \bibinfo {author} {\bibfnamefont {Paolo}\ \bibnamefont {Panci}},
  \ and\ \bibinfo {author} {\bibfnamefont {Alessandro}\ \bibnamefont
  {Strumia}},\ }\bibfield  {title} {\enquote {\bibinfo {title} {{Bounds on Dark
  Matter annihilations from 21 cm data}},}\ }\href {\doibase
  10.1103/PhysRevLett.121.011103} {\bibfield  {journal} {\bibinfo  {journal}
  {Phys. Rev. Lett.}\ }\textbf {\bibinfo {volume} {121}},\ \bibinfo {pages}
  {011103} (\bibinfo {year} {2018})},\ \Eprint
  {http://arxiv.org/abs/1803.03629} {arXiv:1803.03629 [astro-ph.CO]}
  \BibitemShut {NoStop}%
\bibitem [{\citenamefont {Yang}(2018)}]{Yang:2018gjd}%
  \BibitemOpen
  \bibfield  {author} {\bibinfo {author} {\bibfnamefont {Yupeng}\ \bibnamefont
  {Yang}},\ }\bibfield  {title} {\enquote {\bibinfo {title} {{Contributions of
  dark matter annihilation to the global 21 cm spectrum observed by the EDGES
  experiment}},}\ }\href {\doibase 10.1103/PhysRevD.98.103503} {\bibfield
  {journal} {\bibinfo  {journal} {Phys. Rev.}\ }\textbf {\bibinfo {volume}
  {D98}},\ \bibinfo {pages} {103503} (\bibinfo {year} {2018})},\ \Eprint
  {http://arxiv.org/abs/1803.05803} {arXiv:1803.05803 [astro-ph.CO]}
  \BibitemShut {NoStop}%
\bibitem [{\citenamefont {Clark}\ \emph {et~al.}(2018)\citenamefont {Clark},
  \citenamefont {Dutta}, \citenamefont {Gao}, \citenamefont {Ma},\ and\
  \citenamefont {Strigari}}]{Clark:2018ghm}%
  \BibitemOpen
  \bibfield  {author} {\bibinfo {author} {\bibfnamefont {Steven}\ \bibnamefont
  {Clark}}, \bibinfo {author} {\bibfnamefont {Bhaskar}\ \bibnamefont {Dutta}},
  \bibinfo {author} {\bibfnamefont {Yu}~\bibnamefont {Gao}}, \bibinfo {author}
  {\bibfnamefont {Yin-Zhe}\ \bibnamefont {Ma}}, \ and\ \bibinfo {author}
  {\bibfnamefont {Louis~E.}\ \bibnamefont {Strigari}},\ }\bibfield  {title}
  {\enquote {\bibinfo {title} {{21 cm limits on decaying dark matter and
  primordial black holes}},}\ }\href {\doibase 10.1103/PhysRevD.98.043006}
  {\bibfield  {journal} {\bibinfo  {journal} {Phys. Rev.}\ }\textbf {\bibinfo
  {volume} {D98}},\ \bibinfo {pages} {043006} (\bibinfo {year} {2018})},\
  \Eprint {http://arxiv.org/abs/1803.09390} {arXiv:1803.09390 [astro-ph.HE]}
  \BibitemShut {NoStop}%
\bibitem [{\citenamefont {Liu}\ and\ \citenamefont
  {Slatyer}(2018)}]{Liu:2018uzy}%
  \BibitemOpen
  \bibfield  {author} {\bibinfo {author} {\bibfnamefont {Hongwan}\ \bibnamefont
  {Liu}}\ and\ \bibinfo {author} {\bibfnamefont {Tracy~R.}\ \bibnamefont
  {Slatyer}},\ }\bibfield  {title} {\enquote {\bibinfo {title} {{Implications
  of a 21-cm signal for dark matter annihilation and decay}},}\ }\href
  {\doibase 10.1103/PhysRevD.98.023501} {\bibfield  {journal} {\bibinfo
  {journal} {Phys. Rev.}\ }\textbf {\bibinfo {volume} {D98}},\ \bibinfo {pages}
  {023501} (\bibinfo {year} {2018})},\ \Eprint
  {http://arxiv.org/abs/1803.09739} {arXiv:1803.09739 [astro-ph.CO]}
  \BibitemShut {NoStop}%
\bibitem [{\citenamefont {Cheung}\ \emph {et~al.}(2019)\citenamefont {Cheung},
  \citenamefont {Kuo}, \citenamefont {Ng},\ and\ \citenamefont
  {Tsai}}]{Cheung:2018vww}%
  \BibitemOpen
  \bibfield  {author} {\bibinfo {author} {\bibfnamefont {Kingman}\ \bibnamefont
  {Cheung}}, \bibinfo {author} {\bibfnamefont {Jui-Lin}\ \bibnamefont {Kuo}},
  \bibinfo {author} {\bibfnamefont {Kin-Wang}\ \bibnamefont {Ng}}, \ and\
  \bibinfo {author} {\bibfnamefont {Yue-Lin~Sming}\ \bibnamefont {Tsai}},\
  }\bibfield  {title} {\enquote {\bibinfo {title} {{The impact of EDGES 21-cm
  data on dark matter interactions}},}\ }\href {\doibase
  10.1016/j.physletb.2018.11.058} {\bibfield  {journal} {\bibinfo  {journal}
  {Phys. Lett.}\ }\textbf {\bibinfo {volume} {B789}},\ \bibinfo {pages}
  {137--144} (\bibinfo {year} {2019})},\ \Eprint
  {http://arxiv.org/abs/1803.09398} {arXiv:1803.09398 [astro-ph.CO]}
  \BibitemShut {NoStop}%
\bibitem [{\citenamefont {Mitridate}\ and\ \citenamefont
  {Podo}(2018)}]{Mitridate:2018iag}%
  \BibitemOpen
  \bibfield  {author} {\bibinfo {author} {\bibfnamefont {Andrea}\ \bibnamefont
  {Mitridate}}\ and\ \bibinfo {author} {\bibfnamefont {Alessandro}\
  \bibnamefont {Podo}},\ }\bibfield  {title} {\enquote {\bibinfo {title}
  {{Bounds on Dark Matter decay from 21 cm line}},}\ }\href {\doibase
  10.1088/1475-7516/2018/05/069} {\bibfield  {journal} {\bibinfo  {journal}
  {JCAP}\ }\textbf {\bibinfo {volume} {1805}},\ \bibinfo {pages} {069}
  (\bibinfo {year} {2018})},\ \Eprint {http://arxiv.org/abs/1803.11169}
  {arXiv:1803.11169 [hep-ph]} \BibitemShut {NoStop}%
\bibitem [{\citenamefont {Barkana}(2018)}]{Barkana:2018lgd}%
  \BibitemOpen
  \bibfield  {author} {\bibinfo {author} {\bibfnamefont {Rennan}\ \bibnamefont
  {Barkana}},\ }\bibfield  {title} {\enquote {\bibinfo {title} {{Possible
  interaction between baryons and dark-matter particles revealed by the first
  stars}},}\ }\href {\doibase 10.1038/nature25791} {\bibfield  {journal}
  {\bibinfo  {journal} {Nature}\ }\textbf {\bibinfo {volume} {555}},\ \bibinfo
  {pages} {71--74} (\bibinfo {year} {2018})},\ \Eprint
  {http://arxiv.org/abs/1803.06698} {arXiv:1803.06698 [astro-ph.CO]}
  \BibitemShut {NoStop}%
\bibitem [{\citenamefont {Fialkov}\ \emph {et~al.}(2018)\citenamefont
  {Fialkov}, \citenamefont {Barkana},\ and\ \citenamefont
  {Cohen}}]{Fialkov:2018xre}%
  \BibitemOpen
  \bibfield  {author} {\bibinfo {author} {\bibfnamefont {Anastasia}\
  \bibnamefont {Fialkov}}, \bibinfo {author} {\bibfnamefont {Rennan}\
  \bibnamefont {Barkana}}, \ and\ \bibinfo {author} {\bibfnamefont {Aviad}\
  \bibnamefont {Cohen}},\ }\bibfield  {title} {\enquote {\bibinfo {title}
  {{Constraining Baryon--Dark Matter Scattering with the Cosmic Dawn 21-cm
  Signal}},}\ }\href {\doibase 10.1103/PhysRevLett.121.011101} {\bibfield
  {journal} {\bibinfo  {journal} {Phys. Rev. Lett.}\ }\textbf {\bibinfo
  {volume} {121}},\ \bibinfo {pages} {011101} (\bibinfo {year} {2018})},\
  \Eprint {http://arxiv.org/abs/1802.10577} {arXiv:1802.10577 [astro-ph.CO]}
  \BibitemShut {NoStop}%
\bibitem [{\citenamefont {Berlin}\ \emph {et~al.}(2018)\citenamefont {Berlin},
  \citenamefont {Hooper}, \citenamefont {Krnjaic},\ and\ \citenamefont
  {McDermott}}]{Berlin:2018sjs}%
  \BibitemOpen
  \bibfield  {author} {\bibinfo {author} {\bibfnamefont {Asher}\ \bibnamefont
  {Berlin}}, \bibinfo {author} {\bibfnamefont {Dan}\ \bibnamefont {Hooper}},
  \bibinfo {author} {\bibfnamefont {Gordan}\ \bibnamefont {Krnjaic}}, \ and\
  \bibinfo {author} {\bibfnamefont {Samuel~D.}\ \bibnamefont {McDermott}},\
  }\bibfield  {title} {\enquote {\bibinfo {title} {{Severely Constraining Dark
  Matter Interpretations of the 21-cm Anomaly}},}\ }\href {\doibase
  10.1103/PhysRevLett.121.011102} {\bibfield  {journal} {\bibinfo  {journal}
  {Phys. Rev. Lett.}\ }\textbf {\bibinfo {volume} {121}},\ \bibinfo {pages}
  {011102} (\bibinfo {year} {2018})},\ \Eprint
  {http://arxiv.org/abs/1803.02804} {arXiv:1803.02804 [hep-ph]} \BibitemShut
  {NoStop}%
\bibitem [{\citenamefont {Barkana}\ \emph {et~al.}(2018)\citenamefont
  {Barkana}, \citenamefont {Outmezguine}, \citenamefont {Redigolo},\ and\
  \citenamefont {Volansky}}]{Barkana:2018qrx}%
  \BibitemOpen
  \bibfield  {author} {\bibinfo {author} {\bibfnamefont {Rennan}\ \bibnamefont
  {Barkana}}, \bibinfo {author} {\bibfnamefont {Nadav~Joseph}\ \bibnamefont
  {Outmezguine}}, \bibinfo {author} {\bibfnamefont {Diego}\ \bibnamefont
  {Redigolo}}, \ and\ \bibinfo {author} {\bibfnamefont {Tomer}\ \bibnamefont
  {Volansky}},\ }\bibfield  {title} {\enquote {\bibinfo {title} {{Strong
  constraints on light dark matter interpretation of the EDGES signal}},}\
  }\href {\doibase 10.1103/PhysRevD.98.103005} {\bibfield  {journal} {\bibinfo
  {journal} {Phys. Rev.}\ }\textbf {\bibinfo {volume} {D98}},\ \bibinfo {pages}
  {103005} (\bibinfo {year} {2018})},\ \Eprint
  {http://arxiv.org/abs/1803.03091} {arXiv:1803.03091 [hep-ph]} \BibitemShut
  {NoStop}%
\bibitem [{\citenamefont {Fraser}\ \emph {et~al.}(2018)\citenamefont {Fraser}
  \emph {et~al.}}]{Fraser:2018acy}%
  \BibitemOpen
  \bibfield  {author} {\bibinfo {author} {\bibfnamefont {Sean}\ \bibnamefont
  {Fraser}} \emph {et~al.},\ }\bibfield  {title} {\enquote {\bibinfo {title}
  {{The EDGES 21 cm Anomaly and Properties of Dark Matter}},}\ }\href {\doibase
  10.1016/j.physletb.2018.08.035} {\bibfield  {journal} {\bibinfo  {journal}
  {Phys. Lett.}\ }\textbf {\bibinfo {volume} {B785}},\ \bibinfo {pages}
  {159--164} (\bibinfo {year} {2018})},\ \Eprint
  {http://arxiv.org/abs/1803.03245} {arXiv:1803.03245 [hep-ph]} \BibitemShut
  {NoStop}%
\bibitem [{\citenamefont {Slatyer}\ and\ \citenamefont
  {Wu}(2018)}]{Slatyer:2018aqg}%
  \BibitemOpen
  \bibfield  {author} {\bibinfo {author} {\bibfnamefont {Tracy~R.}\
  \bibnamefont {Slatyer}}\ and\ \bibinfo {author} {\bibfnamefont {Chih-Liang}\
  \bibnamefont {Wu}},\ }\bibfield  {title} {\enquote {\bibinfo {title}
  {{Early-Universe constraints on dark matter-baryon scattering and their
  implications for a global 21 cm signal}},}\ }\href {\doibase
  10.1103/PhysRevD.98.023013} {\bibfield  {journal} {\bibinfo  {journal} {Phys.
  Rev.}\ }\textbf {\bibinfo {volume} {D98}},\ \bibinfo {pages} {023013}
  (\bibinfo {year} {2018})},\ \Eprint {http://arxiv.org/abs/1803.09734}
  {arXiv:1803.09734 [astro-ph.CO]} \BibitemShut {NoStop}%
\bibitem [{\citenamefont {Madau}\ \emph {et~al.}(1997)\citenamefont {Madau},
  \citenamefont {Meiksin},\ and\ \citenamefont {Rees}}]{Madau:1996cs}%
  \BibitemOpen
  \bibfield  {author} {\bibinfo {author} {\bibfnamefont {Piero}\ \bibnamefont
  {Madau}}, \bibinfo {author} {\bibfnamefont {Avery}\ \bibnamefont {Meiksin}},
  \ and\ \bibinfo {author} {\bibfnamefont {Martin~J.}\ \bibnamefont {Rees}},\
  }\bibfield  {title} {\enquote {\bibinfo {title} {{21-CM tomography of the
  intergalactic medium at high redshift}},}\ }\href {\doibase 10.1086/303549}
  {\bibfield  {journal} {\bibinfo  {journal} {Astrophys. J.}\ }\textbf
  {\bibinfo {volume} {475}},\ \bibinfo {pages} {429} (\bibinfo {year}
  {1997})},\ \Eprint {http://arxiv.org/abs/astro-ph/9608010}
  {arXiv:astro-ph/9608010 [astro-ph]} \BibitemShut {NoStop}%
\bibitem [{\citenamefont {Furlanetto}\ \emph {et~al.}(2006)\citenamefont
  {Furlanetto}, \citenamefont {Oh},\ and\ \citenamefont
  {Briggs}}]{Furlanetto:2006jb}%
  \BibitemOpen
  \bibfield  {author} {\bibinfo {author} {\bibfnamefont {Steven}\ \bibnamefont
  {Furlanetto}}, \bibinfo {author} {\bibfnamefont {S.~Peng}\ \bibnamefont
  {Oh}}, \ and\ \bibinfo {author} {\bibfnamefont {Frank}\ \bibnamefont
  {Briggs}},\ }\bibfield  {title} {\enquote {\bibinfo {title} {{Cosmology at
  Low Frequencies: The 21 cm Transition and the High-Redshift Universe}},}\
  }\href {\doibase 10.1016/j.physrep.2006.08.002} {\bibfield  {journal}
  {\bibinfo  {journal} {Phys. Rept.}\ }\textbf {\bibinfo {volume} {433}},\
  \bibinfo {pages} {181--301} (\bibinfo {year} {2006})},\ \Eprint
  {http://arxiv.org/abs/astro-ph/0608032} {arXiv:astro-ph/0608032 [astro-ph]}
  \BibitemShut {NoStop}%
\bibitem [{\citenamefont {Zaldarriaga}\ \emph {et~al.}(2004)\citenamefont
  {Zaldarriaga}, \citenamefont {Furlanetto},\ and\ \citenamefont
  {Hernquist}}]{Zaldarriaga:2003du}%
  \BibitemOpen
  \bibfield  {author} {\bibinfo {author} {\bibfnamefont {Matias}\ \bibnamefont
  {Zaldarriaga}}, \bibinfo {author} {\bibfnamefont {Steven~R.}\ \bibnamefont
  {Furlanetto}}, \ and\ \bibinfo {author} {\bibfnamefont {Lars}\ \bibnamefont
  {Hernquist}},\ }\bibfield  {title} {\enquote {\bibinfo {title} {{21
  Centimeter fluctuations from cosmic gas at high redshifts}},}\ }\href
  {\doibase 10.1086/386327} {\bibfield  {journal} {\bibinfo  {journal}
  {Astrophys. J.}\ }\textbf {\bibinfo {volume} {608}},\ \bibinfo {pages}
  {622--635} (\bibinfo {year} {2004})},\ \Eprint
  {http://arxiv.org/abs/astro-ph/0311514} {arXiv:astro-ph/0311514 [astro-ph]}
  \BibitemShut {NoStop}%
\bibitem [{\citenamefont {Ciardi}\ and\ \citenamefont
  {Ferrara}(2005)}]{Ciardi:2004ru}%
  \BibitemOpen
  \bibfield  {author} {\bibinfo {author} {\bibfnamefont {Benedetta}\
  \bibnamefont {Ciardi}}\ and\ \bibinfo {author} {\bibfnamefont {Andrea}\
  \bibnamefont {Ferrara}},\ }\bibfield  {title} {\enquote {\bibinfo {title}
  {{The First cosmic structures and their effects}},}\ }\href {\doibase
  10.1007/s11214-005-3592-0} {\bibfield  {journal} {\bibinfo  {journal} {Space
  Sci. Rev.}\ }\textbf {\bibinfo {volume} {116}},\ \bibinfo {pages} {625--705}
  (\bibinfo {year} {2005})},\ \Eprint {http://arxiv.org/abs/astro-ph/0409018}
  {arXiv:astro-ph/0409018 [astro-ph]} \BibitemShut {NoStop}%
\bibitem [{\citenamefont {Pritchard}\ and\ \citenamefont
  {Loeb}(2012)}]{Pritchard:2011xb}%
  \BibitemOpen
  \bibfield  {author} {\bibinfo {author} {\bibfnamefont {Jonathan~R.}\
  \bibnamefont {Pritchard}}\ and\ \bibinfo {author} {\bibfnamefont {Abraham}\
  \bibnamefont {Loeb}},\ }\bibfield  {title} {\enquote {\bibinfo {title}
  {{21-cm cosmology}},}\ }\href {\doibase 10.1088/0034-4885/75/8/086901}
  {\bibfield  {journal} {\bibinfo  {journal} {Rept. Prog. Phys.}\ }\textbf
  {\bibinfo {volume} {75}},\ \bibinfo {pages} {086901} (\bibinfo {year}
  {2012})},\ \Eprint {http://arxiv.org/abs/1109.6012} {arXiv:1109.6012
  [astro-ph.CO]} \BibitemShut {NoStop}%
\bibitem [{\citenamefont {Barkana}(2016)}]{Barkana:2016nyr}%
  \BibitemOpen
  \bibfield  {author} {\bibinfo {author} {\bibfnamefont {Rennan}\ \bibnamefont
  {Barkana}},\ }\bibfield  {title} {\enquote {\bibinfo {title} {{The Rise of
  the First Stars: Supersonic Streaming, Radiative Feedback, and 21-cm
  Cosmology}},}\ }\href {\doibase 10.1016/j.physrep.2016.06.006} {\bibfield
  {journal} {\bibinfo  {journal} {Phys. Rept.}\ }\textbf {\bibinfo {volume}
  {645}},\ \bibinfo {pages} {1--59} (\bibinfo {year} {2016})},\ \Eprint
  {http://arxiv.org/abs/1605.04357} {arXiv:1605.04357 [astro-ph.CO]}
  \BibitemShut {NoStop}%
\bibitem [{\citenamefont {Pritchard}\ and\ \citenamefont
  {Loeb}(2008)}]{Pritchard:2008da}%
  \BibitemOpen
  \bibfield  {author} {\bibinfo {author} {\bibfnamefont {Jonathan~R.}\
  \bibnamefont {Pritchard}}\ and\ \bibinfo {author} {\bibfnamefont {Abraham}\
  \bibnamefont {Loeb}},\ }\bibfield  {title} {\enquote {\bibinfo {title}
  {{Evolution of the 21 cm signal throughout cosmic history}},}\ }\href
  {\doibase 10.1103/PhysRevD.78.103511} {\bibfield  {journal} {\bibinfo
  {journal} {Phys. Rev.}\ }\textbf {\bibinfo {volume} {D78}},\ \bibinfo {pages}
  {103511} (\bibinfo {year} {2008})},\ \Eprint {http://arxiv.org/abs/0802.2102}
  {arXiv:0802.2102 [astro-ph]} \BibitemShut {NoStop}%
\bibitem [{\citenamefont {Mirocha}(2014)}]{Mirocha:2014faa}%
  \BibitemOpen
  \bibfield  {author} {\bibinfo {author} {\bibfnamefont {Jordan}\ \bibnamefont
  {Mirocha}},\ }\bibfield  {title} {\enquote {\bibinfo {title} {{Decoding the
  X-ray Properties of Pre-Reionization Era Sources}},}\ }\href {\doibase
  10.1093/mnras/stu1193} {\bibfield  {journal} {\bibinfo  {journal} {Mon. Not.
  Roy. Astron. Soc.}\ }\textbf {\bibinfo {volume} {443}},\ \bibinfo {pages}
  {1211--1223} (\bibinfo {year} {2014})},\ \Eprint
  {http://arxiv.org/abs/1406.4120} {arXiv:1406.4120 [astro-ph.GA]} \BibitemShut
  {NoStop}%
\bibitem [{\citenamefont {Mesinger}\ \emph {et~al.}(2011)\citenamefont
  {Mesinger}, \citenamefont {Furlanetto},\ and\ \citenamefont
  {Cen}}]{Mesinger:2010ne}%
  \BibitemOpen
  \bibfield  {author} {\bibinfo {author} {\bibfnamefont {Andrei}\ \bibnamefont
  {Mesinger}}, \bibinfo {author} {\bibfnamefont {Steven}\ \bibnamefont
  {Furlanetto}}, \ and\ \bibinfo {author} {\bibfnamefont {Renyue}\ \bibnamefont
  {Cen}},\ }\bibfield  {title} {\enquote {\bibinfo {title} {{21cmFAST: A Fast,
  Semi-Numerical Simulation of the High-Redshift 21-cm Signal}},}\ }\href
  {\doibase 10.1111/j.1365-2966.2010.17731.x} {\bibfield  {journal} {\bibinfo
  {journal} {Mon. Not. Roy. Astron. Soc.}\ }\textbf {\bibinfo {volume} {411}},\
  \bibinfo {pages} {955} (\bibinfo {year} {2011})},\ \Eprint
  {http://arxiv.org/abs/1003.3878} {arXiv:1003.3878 [astro-ph.CO]} \BibitemShut
  {NoStop}%
\bibitem [{\citenamefont {Bond}\ \emph {et~al.}(1980)\citenamefont {Bond},
  \citenamefont {Efstathiou},\ and\ \citenamefont {Silk}}]{Bond:1980ha}%
  \BibitemOpen
  \bibfield  {author} {\bibinfo {author} {\bibfnamefont {J.~R.}\ \bibnamefont
  {Bond}}, \bibinfo {author} {\bibfnamefont {G.}~\bibnamefont {Efstathiou}}, \
  and\ \bibinfo {author} {\bibfnamefont {J.}~\bibnamefont {Silk}},\ }\bibfield
  {title} {\enquote {\bibinfo {title} {{Massive Neutrinos and the Large Scale
  Structure of the Universe}},}\ }\href {\doibase 10.1103/PhysRevLett.45.1980}
  {\bibfield  {journal} {\bibinfo  {journal} {Phys. Rev. Lett.}\ }\textbf
  {\bibinfo {volume} {45}},\ \bibinfo {pages} {1980--1984} (\bibinfo {year}
  {1980})}\BibitemShut {NoStop}%
\bibitem [{\citenamefont {Boyarsky}\ \emph
  {et~al.}(2009{\natexlab{a}})\citenamefont {Boyarsky}, \citenamefont
  {Lesgourgues}, \citenamefont {Ruchayskiy},\ and\ \citenamefont
  {Viel}}]{Boyarsky:2008xj}%
  \BibitemOpen
  \bibfield  {author} {\bibinfo {author} {\bibfnamefont {Alexey}\ \bibnamefont
  {Boyarsky}}, \bibinfo {author} {\bibfnamefont {Julien}\ \bibnamefont
  {Lesgourgues}}, \bibinfo {author} {\bibfnamefont {Oleg}\ \bibnamefont
  {Ruchayskiy}}, \ and\ \bibinfo {author} {\bibfnamefont {Matteo}\ \bibnamefont
  {Viel}},\ }\bibfield  {title} {\enquote {\bibinfo {title} {{Lyman-alpha
  constraints on warm and on warm-plus-cold dark matter models}},}\ }\href
  {\doibase 10.1088/1475-7516/2009/05/012} {\bibfield  {journal} {\bibinfo
  {journal} {JCAP}\ }\textbf {\bibinfo {volume} {0905}},\ \bibinfo {pages}
  {012} (\bibinfo {year} {2009}{\natexlab{a}})},\ \Eprint
  {http://arxiv.org/abs/0812.0010} {arXiv:0812.0010 [astro-ph]} \BibitemShut
  {NoStop}%
\bibitem [{\citenamefont {Boyarsky}\ \emph {et~al.}(2019)\citenamefont
  {Boyarsky}, \citenamefont {Drewes}, \citenamefont {Lasserre}, \citenamefont
  {Mertens},\ and\ \citenamefont {Ruchayskiy}}]{Boyarsky:2018tvu}%
  \BibitemOpen
  \bibfield  {author} {\bibinfo {author} {\bibfnamefont {A.}~\bibnamefont
  {Boyarsky}}, \bibinfo {author} {\bibfnamefont {M.}~\bibnamefont {Drewes}},
  \bibinfo {author} {\bibfnamefont {T.}~\bibnamefont {Lasserre}}, \bibinfo
  {author} {\bibfnamefont {S.}~\bibnamefont {Mertens}}, \ and\ \bibinfo
  {author} {\bibfnamefont {O.}~\bibnamefont {Ruchayskiy}},\ }\bibfield  {title}
  {\enquote {\bibinfo {title} {{Sterile Neutrino Dark Matter}},}\ }\href
  {\doibase 10.1016/j.ppnp.2018.07.004} {\bibfield  {journal} {\bibinfo
  {journal} {Prog. Part. Nucl. Phys.}\ }\textbf {\bibinfo {volume} {104}},\
  \bibinfo {pages} {1--45} (\bibinfo {year} {2019})},\ \Eprint
  {http://arxiv.org/abs/1807.07938} {arXiv:1807.07938 [hep-ph]} \BibitemShut
  {NoStop}%
\bibitem [{\citenamefont {Dodelson}\ and\ \citenamefont
  {Widrow}(1994)}]{Dodelson:1993je}%
  \BibitemOpen
  \bibfield  {author} {\bibinfo {author} {\bibfnamefont {Scott}\ \bibnamefont
  {Dodelson}}\ and\ \bibinfo {author} {\bibfnamefont {Lawrence~M.}\
  \bibnamefont {Widrow}},\ }\bibfield  {title} {\enquote {\bibinfo {title}
  {{Sterile-neutrinos as dark matter}},}\ }\href {\doibase
  10.1103/PhysRevLett.72.17} {\bibfield  {journal} {\bibinfo  {journal}
  {Phys.Rev.Lett.}\ }\textbf {\bibinfo {volume} {72}},\ \bibinfo {pages}
  {17--20} (\bibinfo {year} {1994})}\BibitemShut {NoStop}%
\bibitem [{\citenamefont {Viel}\ \emph {et~al.}(2005)\citenamefont {Viel},
  \citenamefont {Lesgourgues}, \citenamefont {Haehnelt}, \citenamefont
  {Matarrese},\ and\ \citenamefont {Riotto}}]{Viel:2005qj}%
  \BibitemOpen
  \bibfield  {author} {\bibinfo {author} {\bibfnamefont {Matteo}\ \bibnamefont
  {Viel}}, \bibinfo {author} {\bibfnamefont {Julien}\ \bibnamefont
  {Lesgourgues}}, \bibinfo {author} {\bibfnamefont {Martin~G.}\ \bibnamefont
  {Haehnelt}}, \bibinfo {author} {\bibfnamefont {Sabino}\ \bibnamefont
  {Matarrese}}, \ and\ \bibinfo {author} {\bibfnamefont {Antonio}\ \bibnamefont
  {Riotto}},\ }\bibfield  {title} {\enquote {\bibinfo {title} {{Constraining
  warm dark matter candidates including sterile neutrinos and light gravitinos
  with WMAP and the Lyman-alpha forest}},}\ }\href {\doibase
  10.1103/PhysRevD.71.063534} {\bibfield  {journal} {\bibinfo  {journal} {Phys.
  Rev.}\ }\textbf {\bibinfo {volume} {D71}},\ \bibinfo {pages} {063534}
  (\bibinfo {year} {2005})}\BibitemShut {NoStop}%
\bibitem [{\citenamefont {Benson}\ \emph {et~al.}(2013)\citenamefont {Benson},
  \citenamefont {Farahi}, \citenamefont {Cole}, \citenamefont {Moustakas},
  \citenamefont {Jenkins}, \citenamefont {Lovell}, \citenamefont {Kennedy},
  \citenamefont {Helly},\ and\ \citenamefont {Frenk}}]{Benson:2012su}%
  \BibitemOpen
  \bibfield  {author} {\bibinfo {author} {\bibfnamefont {Andrew~J.}\
  \bibnamefont {Benson}}, \bibinfo {author} {\bibfnamefont {Arya}\ \bibnamefont
  {Farahi}}, \bibinfo {author} {\bibfnamefont {Shaun}\ \bibnamefont {Cole}},
  \bibinfo {author} {\bibfnamefont {Leonidas~A.}\ \bibnamefont {Moustakas}},
  \bibinfo {author} {\bibfnamefont {Adrian}\ \bibnamefont {Jenkins}}, \bibinfo
  {author} {\bibfnamefont {Mark}\ \bibnamefont {Lovell}}, \bibinfo {author}
  {\bibfnamefont {Rachel}\ \bibnamefont {Kennedy}}, \bibinfo {author}
  {\bibfnamefont {John}\ \bibnamefont {Helly}}, \ and\ \bibinfo {author}
  {\bibfnamefont {Carlos}\ \bibnamefont {Frenk}},\ }\bibfield  {title}
  {\enquote {\bibinfo {title} {{Dark Matter Halo Merger Histories Beyond Cold
  Dark Matter: I - Methods and Application to Warm Dark Matter}},}\ }\href
  {\doibase 10.1093/mnras/sts159} {\bibfield  {journal} {\bibinfo  {journal}
  {Mon. Not. Roy. Astron. Soc.}\ }\textbf {\bibinfo {volume} {428}},\ \bibinfo
  {pages} {1774} (\bibinfo {year} {2013})},\ \Eprint
  {http://arxiv.org/abs/1209.3018} {arXiv:1209.3018 [astro-ph.CO]} \BibitemShut
  {NoStop}%
\bibitem [{\citenamefont {Schneider}(2018)}]{Schneider:2018xba}%
  \BibitemOpen
  \bibfield  {author} {\bibinfo {author} {\bibfnamefont {Aurel}\ \bibnamefont
  {Schneider}},\ }\bibfield  {title} {\enquote {\bibinfo {title} {{Constraining
  noncold dark matter models with the global 21-cm signal}},}\ }\href {\doibase
  10.1103/PhysRevD.98.063021} {\bibfield  {journal} {\bibinfo  {journal} {Phys.
  Rev.}\ }\textbf {\bibinfo {volume} {D98}},\ \bibinfo {pages} {063021}
  (\bibinfo {year} {2018})},\ \Eprint {http://arxiv.org/abs/1805.00021}
  {arXiv:1805.00021 [astro-ph.CO]} \BibitemShut {NoStop}%
\bibitem [{\citenamefont {Haiman}\ \emph {et~al.}(2000)\citenamefont {Haiman},
  \citenamefont {Abel},\ and\ \citenamefont {Rees}}]{Haiman:1999mn}%
  \BibitemOpen
  \bibfield  {author} {\bibinfo {author} {\bibfnamefont {Zoltan}\ \bibnamefont
  {Haiman}}, \bibinfo {author} {\bibfnamefont {Tom}\ \bibnamefont {Abel}}, \
  and\ \bibinfo {author} {\bibfnamefont {Martin~J.}\ \bibnamefont {Rees}},\
  }\bibfield  {title} {\enquote {\bibinfo {title} {{The radiative feedback of
  the first cosmological objects}},}\ }\href {\doibase 10.1086/308723}
  {\bibfield  {journal} {\bibinfo  {journal} {Astrophys. J.}\ }\textbf
  {\bibinfo {volume} {534}},\ \bibinfo {pages} {11--24} (\bibinfo {year}
  {2000})},\ \Eprint {http://arxiv.org/abs/astro-ph/9903336}
  {arXiv:astro-ph/9903336 [astro-ph]} \BibitemShut {NoStop}%
\bibitem [{\citenamefont {Barkana}\ and\ \citenamefont
  {Loeb}(2001)}]{Barkana:2000fd}%
  \BibitemOpen
  \bibfield  {author} {\bibinfo {author} {\bibfnamefont {Rennan}\ \bibnamefont
  {Barkana}}\ and\ \bibinfo {author} {\bibfnamefont {Abraham}\ \bibnamefont
  {Loeb}},\ }\bibfield  {title} {\enquote {\bibinfo {title} {{In the beginning:
  The First sources of light and the reionization of the Universe}},}\ }\href
  {\doibase 10.1016/S0370-1573(01)00019-9} {\bibfield  {journal} {\bibinfo
  {journal} {Phys. Rept.}\ }\textbf {\bibinfo {volume} {349}},\ \bibinfo
  {pages} {125--238} (\bibinfo {year} {2001})},\ \Eprint
  {http://arxiv.org/abs/astro-ph/0010468} {arXiv:astro-ph/0010468 [astro-ph]}
  \BibitemShut {NoStop}%
\bibitem [{\citenamefont {Gao}\ and\ \citenamefont
  {Theuns}(2007)}]{Gao:2007yk}%
  \BibitemOpen
  \bibfield  {author} {\bibinfo {author} {\bibfnamefont {Liang}\ \bibnamefont
  {Gao}}\ and\ \bibinfo {author} {\bibfnamefont {Tom}\ \bibnamefont {Theuns}},\
  }\bibfield  {title} {\enquote {\bibinfo {title} {{Lighting the Universe with
  filaments}},}\ }\href {\doibase 10.1126/science.1146676} {\bibfield
  {journal} {\bibinfo  {journal} {Science}\ }\textbf {\bibinfo {volume}
  {317}},\ \bibinfo {pages} {1527} (\bibinfo {year} {2007})},\ \Eprint
  {http://arxiv.org/abs/0709.2165} {arXiv:0709.2165 [astro-ph]} \BibitemShut
  {NoStop}%
\bibitem [{\citenamefont {Paduroiu}\ \emph {et~al.}(2015)\citenamefont
  {Paduroiu}, \citenamefont {Revaz},\ and\ \citenamefont
  {Pfenniger}}]{Paduroiu:2015jfa}%
  \BibitemOpen
  \bibfield  {author} {\bibinfo {author} {\bibfnamefont {Sinziana}\
  \bibnamefont {Paduroiu}}, \bibinfo {author} {\bibfnamefont {Yves}\
  \bibnamefont {Revaz}}, \ and\ \bibinfo {author} {\bibfnamefont {Daniel}\
  \bibnamefont {Pfenniger}},\ }\bibfield  {title} {\enquote {\bibinfo {title}
  {{Structure formation in warm dark matter cosmologies: Top-Bottom
  Upside-Down}},}\ }\href@noop {} {\  (\bibinfo {year} {2015})},\ \Eprint
  {http://arxiv.org/abs/1506.03789} {arXiv:1506.03789 [astro-ph.CO]}
  \BibitemShut {NoStop}%
\bibitem [{\citenamefont {Leo}\ \emph {et~al.}(2019)\citenamefont {Leo},
  \citenamefont {Theuns}, \citenamefont {Baugh}, \citenamefont {Li},\ and\
  \citenamefont {Pascoli}}]{Leo:2019gwh}%
  \BibitemOpen
  \bibfield  {author} {\bibinfo {author} {\bibfnamefont {Matteo}\ \bibnamefont
  {Leo}}, \bibinfo {author} {\bibfnamefont {Tom}\ \bibnamefont {Theuns}},
  \bibinfo {author} {\bibfnamefont {Carlton~M.}\ \bibnamefont {Baugh}},
  \bibinfo {author} {\bibfnamefont {Baojiu}\ \bibnamefont {Li}}, \ and\
  \bibinfo {author} {\bibfnamefont {Silvia}\ \bibnamefont {Pascoli}},\
  }\bibfield  {title} {\enquote {\bibinfo {title} {{Constraining structure
  formation using EDGES}},}\ }\href@noop {} {\  (\bibinfo {year} {2019})},\
  \Eprint {http://arxiv.org/abs/1909.04641} {arXiv:1909.04641 [astro-ph.CO]}
  \BibitemShut {NoStop}%
\bibitem [{\citenamefont {Chatterjee}\ \emph {et~al.}(2019)\citenamefont
  {Chatterjee}, \citenamefont {Dayal}, \citenamefont {Choudhury},\ and\
  \citenamefont {Hutter}}]{Chatterjee:2019jts}%
  \BibitemOpen
  \bibfield  {author} {\bibinfo {author} {\bibfnamefont {Atrideb}\ \bibnamefont
  {Chatterjee}}, \bibinfo {author} {\bibfnamefont {Pratika}\ \bibnamefont
  {Dayal}}, \bibinfo {author} {\bibfnamefont {Tirthankar~Roy}\ \bibnamefont
  {Choudhury}}, \ and\ \bibinfo {author} {\bibfnamefont {Anne}\ \bibnamefont
  {Hutter}},\ }\bibfield  {title} {\enquote {\bibinfo {title} {{Ruling out 3
  keV warm dark matter using 21 cm EDGES data}},}\ }\href {\doibase
  10.1093/mnras/stz1444} {\bibfield  {journal} {\bibinfo  {journal} {Mon. Not.
  Roy. Astron. Soc.}\ }\textbf {\bibinfo {volume} {487}},\ \bibinfo {pages}
  {3560--3567} (\bibinfo {year} {2019})},\ \Eprint
  {http://arxiv.org/abs/1902.09562} {arXiv:1902.09562 [astro-ph.CO]}
  \BibitemShut {NoStop}%
\bibitem [{\citenamefont {Lopez-Honorez}\ \emph {et~al.}(2019)\citenamefont
  {Lopez-Honorez}, \citenamefont {Mena},\ and\ \citenamefont
  {Villanueva-Domingo}}]{Lopez-Honorez:2018ipk}%
  \BibitemOpen
  \bibfield  {author} {\bibinfo {author} {\bibfnamefont {Laura}\ \bibnamefont
  {Lopez-Honorez}}, \bibinfo {author} {\bibfnamefont {Olga}\ \bibnamefont
  {Mena}}, \ and\ \bibinfo {author} {\bibfnamefont {Pablo}\ \bibnamefont
  {Villanueva-Domingo}},\ }\bibfield  {title} {\enquote {\bibinfo {title}
  {{Dark matter microphysics and 21 cm observations}},}\ }\href {\doibase
  10.1103/PhysRevD.99.023522} {\bibfield  {journal} {\bibinfo  {journal} {Phys.
  Rev.}\ }\textbf {\bibinfo {volume} {D99}},\ \bibinfo {pages} {023522}
  (\bibinfo {year} {2019})},\ \Eprint {http://arxiv.org/abs/1811.02716}
  {arXiv:1811.02716 [astro-ph.CO]} \BibitemShut {NoStop}%
\bibitem [{\citenamefont {Gao}\ \emph {et~al.}(2015)\citenamefont {Gao},
  \citenamefont {Theuns},\ and\ \citenamefont {Springel}}]{Gao:2014yja}%
  \BibitemOpen
  \bibfield  {author} {\bibinfo {author} {\bibfnamefont {Liang}\ \bibnamefont
  {Gao}}, \bibinfo {author} {\bibfnamefont {Tom}\ \bibnamefont {Theuns}}, \
  and\ \bibinfo {author} {\bibfnamefont {Volker}\ \bibnamefont {Springel}},\
  }\bibfield  {title} {\enquote {\bibinfo {title} {{Star forming filaments in
  warm dark matter models}},}\ }\href {\doibase 10.1093/mnras/stv643}
  {\bibfield  {journal} {\bibinfo  {journal} {Mon. Not. Roy. Astron. Soc.}\
  }\textbf {\bibinfo {volume} {450}},\ \bibinfo {pages} {45--52} (\bibinfo
  {year} {2015})},\ \Eprint {http://arxiv.org/abs/1403.2475} {arXiv:1403.2475
  [astro-ph.CO]} \BibitemShut {NoStop}%
\bibitem [{\citenamefont {Herpich}\ \emph {et~al.}(2014)\citenamefont
  {Herpich}, \citenamefont {Stinson}, \citenamefont {Macciò}, \citenamefont
  {Brook}, \citenamefont {Wadsley}, \citenamefont {Couchman},\ and\
  \citenamefont {Quinn}}]{Herpich:2013yga}%
  \BibitemOpen
  \bibfield  {author} {\bibinfo {author} {\bibfnamefont {Jakob}\ \bibnamefont
  {Herpich}}, \bibinfo {author} {\bibfnamefont {Gregory~S.}\ \bibnamefont
  {Stinson}}, \bibinfo {author} {\bibfnamefont {Andrea~V.}\ \bibnamefont
  {Macciò}}, \bibinfo {author} {\bibfnamefont {Chris}\ \bibnamefont {Brook}},
  \bibinfo {author} {\bibfnamefont {James}\ \bibnamefont {Wadsley}}, \bibinfo
  {author} {\bibfnamefont {Hugh M.~P.}\ \bibnamefont {Couchman}}, \ and\
  \bibinfo {author} {\bibfnamefont {Tom}\ \bibnamefont {Quinn}},\ }\bibfield
  {title} {\enquote {\bibinfo {title} {{MaGICC-WDM: the effects of warm dark
  matter in hydrodynamical simulations of disc galaxy formation}},}\ }\href
  {\doibase 10.1093/mnras/stt1883} {\bibfield  {journal} {\bibinfo  {journal}
  {Mon. Not. Roy. Astron. Soc.}\ }\textbf {\bibinfo {volume} {437}},\ \bibinfo
  {pages} {293--304} (\bibinfo {year} {2014})},\ \Eprint
  {http://arxiv.org/abs/1308.1088} {arXiv:1308.1088 [astro-ph.CO]} \BibitemShut
  {NoStop}%
\bibitem [{\citenamefont {Maio}\ and\ \citenamefont
  {Viel}(2015)}]{Maio:2014qwa}%
  \BibitemOpen
  \bibfield  {author} {\bibinfo {author} {\bibfnamefont {Umberto}\ \bibnamefont
  {Maio}}\ and\ \bibinfo {author} {\bibfnamefont {Matteo}\ \bibnamefont
  {Viel}},\ }\bibfield  {title} {\enquote {\bibinfo {title} {{The First Billion
  Years of a Warm Dark Matter Universe}},}\ }\href {\doibase
  10.1093/mnras/stu2304} {\bibfield  {journal} {\bibinfo  {journal} {Mon. Not.
  Roy. Astron. Soc.}\ }\textbf {\bibinfo {volume} {446}},\ \bibinfo {pages}
  {2760--2775} (\bibinfo {year} {2015})},\ \Eprint
  {http://arxiv.org/abs/1409.6718} {arXiv:1409.6718 [astro-ph.CO]} \BibitemShut
  {NoStop}%
\bibitem [{\citenamefont {Colin}\ \emph {et~al.}(2015)\citenamefont {Colin},
  \citenamefont {Avila-Reese}, \citenamefont {Gonzalez-Samaniego},\ and\
  \citenamefont {Velazquez}}]{Colin:2014sga}%
  \BibitemOpen
  \bibfield  {author} {\bibinfo {author} {\bibfnamefont {Pedro}\ \bibnamefont
  {Colin}}, \bibinfo {author} {\bibfnamefont {Vladimir}\ \bibnamefont
  {Avila-Reese}}, \bibinfo {author} {\bibfnamefont {Alejandro}\ \bibnamefont
  {Gonzalez-Samaniego}}, \ and\ \bibinfo {author} {\bibfnamefont {Hector}\
  \bibnamefont {Velazquez}},\ }\bibfield  {title} {\enquote {\bibinfo {title}
  {{Simulations of galaxies formed in warm dark matter halos of masses at the
  filtering scale}},}\ }\href {\doibase 10.1088/0004-637X/803/1/28} {\bibfield
  {journal} {\bibinfo  {journal} {Astrophys. J.}\ }\textbf {\bibinfo {volume}
  {803}},\ \bibinfo {pages} {28} (\bibinfo {year} {2015})},\ \Eprint
  {http://arxiv.org/abs/1412.1100} {arXiv:1412.1100 [astro-ph.GA]} \BibitemShut
  {NoStop}%
\bibitem [{\citenamefont {Power}\ and\ \citenamefont
  {Robotham}(2016)}]{Power:2016ach}%
  \BibitemOpen
  \bibfield  {author} {\bibinfo {author} {\bibfnamefont {Chris}\ \bibnamefont
  {Power}}\ and\ \bibinfo {author} {\bibfnamefont {Aaron S.~G.}\ \bibnamefont
  {Robotham}},\ }\bibfield  {title} {\enquote {\bibinfo {title} {{The Extended
  Stellar Component of Galaxies the Nature of Dark Matter}},}\ }\href {\doibase
  10.3847/0004-637X/825/1/31} {\bibfield  {journal} {\bibinfo  {journal}
  {Astrophys. J.}\ }\textbf {\bibinfo {volume} {825}},\ \bibinfo {pages} {31}
  (\bibinfo {year} {2016})},\ \Eprint {http://arxiv.org/abs/1603.07422}
  {arXiv:1603.07422 [astro-ph.GA]} \BibitemShut {NoStop}%
\bibitem [{\citenamefont {Lovell}\ \emph
  {et~al.}(2017{\natexlab{a}})\citenamefont {Lovell} \emph
  {et~al.}}]{Lovell:2016fec}%
  \BibitemOpen
  \bibfield  {author} {\bibinfo {author} {\bibfnamefont {Mark~R.}\ \bibnamefont
  {Lovell}} \emph {et~al.},\ }\bibfield  {title} {\enquote {\bibinfo {title}
  {{Properties of Local Group galaxies in hydrodynamical simulations of sterile
  neutrino dark matter cosmologies}},}\ }\href {\doibase 10.1093/mnras/stx654}
  {\bibfield  {journal} {\bibinfo  {journal} {Mon. Not. Roy. Astron. Soc.}\
  }\textbf {\bibinfo {volume} {468}},\ \bibinfo {pages} {4285--4298} (\bibinfo
  {year} {2017}{\natexlab{a}})},\ \Eprint {http://arxiv.org/abs/1611.00010}
  {arXiv:1611.00010 [astro-ph.GA]} \BibitemShut {NoStop}%
\bibitem [{\citenamefont {Menci}\ \emph {et~al.}(2012)\citenamefont {Menci},
  \citenamefont {Fiore},\ and\ \citenamefont {Lamastra}}]{Menci:2012kk}%
  \BibitemOpen
  \bibfield  {author} {\bibinfo {author} {\bibfnamefont {N.}~\bibnamefont
  {Menci}}, \bibinfo {author} {\bibfnamefont {F.}~\bibnamefont {Fiore}}, \ and\
  \bibinfo {author} {\bibfnamefont {A.}~\bibnamefont {Lamastra}},\ }\bibfield
  {title} {\enquote {\bibinfo {title} {{Galaxy Formation in WDM Cosmology}},}\
  }\href {\doibase 10.1111/j.1365-2966.2012.20470.x} {\bibfield  {journal}
  {\bibinfo  {journal} {Mon. Not. Roy. Astron. Soc.}\ }\textbf {\bibinfo
  {volume} {421}},\ \bibinfo {pages} {2384} (\bibinfo {year} {2012})},\ \Eprint
  {http://arxiv.org/abs/1201.1617} {arXiv:1201.1617 [astro-ph.CO]} \BibitemShut
  {NoStop}%
\bibitem [{\citenamefont {Kang}\ \emph {et~al.}(2013)\citenamefont {Kang},
  \citenamefont {Maccio},\ and\ \citenamefont {Dutton}}]{Kang:2012up}%
  \BibitemOpen
  \bibfield  {author} {\bibinfo {author} {\bibfnamefont {Xi}~\bibnamefont
  {Kang}}, \bibinfo {author} {\bibfnamefont {Andrea~V.}\ \bibnamefont
  {Maccio}}, \ and\ \bibinfo {author} {\bibfnamefont {Aaron~A.}\ \bibnamefont
  {Dutton}},\ }\bibfield  {title} {\enquote {\bibinfo {title} {{The effect of
  Warm Dark Matter on galaxy properties: constraints from the stellar mass
  function and the Tully-Fisher relation}},}\ }\href {\doibase
  10.1088/0004-637X/767/1/22} {\bibfield  {journal} {\bibinfo  {journal}
  {Astrophys. J.}\ }\textbf {\bibinfo {volume} {767}},\ \bibinfo {pages} {22}
  (\bibinfo {year} {2013})},\ \Eprint {http://arxiv.org/abs/1208.0008}
  {arXiv:1208.0008 [astro-ph.CO]} \BibitemShut {NoStop}%
\bibitem [{\citenamefont {Menci}\ \emph {et~al.}(2013)\citenamefont {Menci},
  \citenamefont {Fiore},\ and\ \citenamefont {Lamastra}}]{Menci:2013ght}%
  \BibitemOpen
  \bibfield  {author} {\bibinfo {author} {\bibfnamefont {N.}~\bibnamefont
  {Menci}}, \bibinfo {author} {\bibfnamefont {F.}~\bibnamefont {Fiore}}, \ and\
  \bibinfo {author} {\bibfnamefont {A.}~\bibnamefont {Lamastra}},\ }\bibfield
  {title} {\enquote {\bibinfo {title} {{The Evolution of Active Galactic Nuclei
  in Warm Dark Matter Cosmology}},}\ }\href {\doibase
  10.1088/0004-637X/766/2/110} {\bibfield  {journal} {\bibinfo  {journal}
  {Astrophys. J.}\ }\textbf {\bibinfo {volume} {766}},\ \bibinfo {pages} {110}
  (\bibinfo {year} {2013})},\ \Eprint {http://arxiv.org/abs/1302.2000}
  {arXiv:1302.2000 [astro-ph.CO]} \BibitemShut {NoStop}%
\bibitem [{\citenamefont {Nierenberg}\ \emph {et~al.}(2013)\citenamefont
  {Nierenberg}, \citenamefont {Treu}, \citenamefont {Menci}, \citenamefont
  {Lu},\ and\ \citenamefont {Wang}}]{Nierenberg:2013lqa}%
  \BibitemOpen
  \bibfield  {author} {\bibinfo {author} {\bibfnamefont {A.~M.}\ \bibnamefont
  {Nierenberg}}, \bibinfo {author} {\bibfnamefont {T.}~\bibnamefont {Treu}},
  \bibinfo {author} {\bibfnamefont {N.}~\bibnamefont {Menci}}, \bibinfo
  {author} {\bibfnamefont {Y.}~\bibnamefont {Lu}}, \ and\ \bibinfo {author}
  {\bibfnamefont {W.}~\bibnamefont {Wang}},\ }\bibfield  {title} {\enquote
  {\bibinfo {title} {{The Cosmic Evolution of Faint Satellite Galaxies as a
  Test of Galaxy Formation and the Nature of Dark Matter}},}\ }\href {\doibase
  10.1088/0004-637X/772/2/146} {\bibfield  {journal} {\bibinfo  {journal}
  {Astrophys. J.}\ }\textbf {\bibinfo {volume} {772}},\ \bibinfo {pages} {146}
  (\bibinfo {year} {2013})},\ \Eprint {http://arxiv.org/abs/1302.3243}
  {arXiv:1302.3243 [astro-ph.CO]} \BibitemShut {NoStop}%
\bibitem [{\citenamefont {Lovell}\ \emph {et~al.}(2016)\citenamefont {Lovell},
  \citenamefont {Bose}, \citenamefont {Boyarsky}, \citenamefont {Cole},
  \citenamefont {Frenk}, \citenamefont {Gonzalez-Perez}, \citenamefont
  {Kennedy}, \citenamefont {Ruchayskiy},\ and\ \citenamefont
  {Smith}}]{Lovell:2015psz}%
  \BibitemOpen
  \bibfield  {author} {\bibinfo {author} {\bibfnamefont {Mark~R.}\ \bibnamefont
  {Lovell}}, \bibinfo {author} {\bibfnamefont {Sownak}\ \bibnamefont {Bose}},
  \bibinfo {author} {\bibfnamefont {Alexey}\ \bibnamefont {Boyarsky}}, \bibinfo
  {author} {\bibfnamefont {Shaun}\ \bibnamefont {Cole}}, \bibinfo {author}
  {\bibfnamefont {Carlos~S.}\ \bibnamefont {Frenk}}, \bibinfo {author}
  {\bibfnamefont {Violeta}\ \bibnamefont {Gonzalez-Perez}}, \bibinfo {author}
  {\bibfnamefont {Rachel}\ \bibnamefont {Kennedy}}, \bibinfo {author}
  {\bibfnamefont {Oleg}\ \bibnamefont {Ruchayskiy}}, \ and\ \bibinfo {author}
  {\bibfnamefont {Alex}\ \bibnamefont {Smith}},\ }\bibfield  {title} {\enquote
  {\bibinfo {title} {{Satellite galaxies in semi-analytic models of galaxy
  formation with sterile neutrino dark matter}},}\ }\href {\doibase
  10.1093/mnras/stw1317} {\bibfield  {journal} {\bibinfo  {journal} {Mon. Not.
  Roy. Astron. Soc.}\ }\textbf {\bibinfo {volume} {461}},\ \bibinfo {pages}
  {60--72} (\bibinfo {year} {2016})},\ \Eprint
  {http://arxiv.org/abs/1511.04078} {arXiv:1511.04078 [astro-ph.CO]}
  \BibitemShut {NoStop}%
\bibitem [{\citenamefont {Lovell}\ \emph
  {et~al.}(2017{\natexlab{b}})\citenamefont {Lovell}, \citenamefont
  {Gonzalez-Perez}, \citenamefont {Bose}, \citenamefont {Boyarsky},
  \citenamefont {Cole}, \citenamefont {Frenk},\ and\ \citenamefont
  {Ruchayskiy}}]{Lovell:2016nkp}%
  \BibitemOpen
  \bibfield  {author} {\bibinfo {author} {\bibfnamefont {Mark~R.}\ \bibnamefont
  {Lovell}}, \bibinfo {author} {\bibfnamefont {Violeta}\ \bibnamefont
  {Gonzalez-Perez}}, \bibinfo {author} {\bibfnamefont {Sownak}\ \bibnamefont
  {Bose}}, \bibinfo {author} {\bibfnamefont {Alexey}\ \bibnamefont {Boyarsky}},
  \bibinfo {author} {\bibfnamefont {Shaun}\ \bibnamefont {Cole}}, \bibinfo
  {author} {\bibfnamefont {Carlos~S.}\ \bibnamefont {Frenk}}, \ and\ \bibinfo
  {author} {\bibfnamefont {Oleg}\ \bibnamefont {Ruchayskiy}},\ }\bibfield
  {title} {\enquote {\bibinfo {title} {{Addressing the too big to fail problem
  with baryon physics and sterile neutrino dark matter}},}\ }\href {\doibase
  10.1093/mnras/stx621} {\bibfield  {journal} {\bibinfo  {journal} {Mon. Not.
  Roy. Astron. Soc.}\ }\textbf {\bibinfo {volume} {468}},\ \bibinfo {pages}
  {2836--2849} (\bibinfo {year} {2017}{\natexlab{b}})},\ \Eprint
  {http://arxiv.org/abs/1611.00005} {arXiv:1611.00005 [astro-ph.GA]}
  \BibitemShut {NoStop}%
\bibitem [{\citenamefont {Wang}\ \emph {et~al.}(2017)\citenamefont {Wang},
  \citenamefont {Gonzalez-Perez}, \citenamefont {Xie}, \citenamefont {Cooper},
  \citenamefont {Frenk}, \citenamefont {Gao}, \citenamefont {Hellwing},
  \citenamefont {Helly}, \citenamefont {Lovell},\ and\ \citenamefont
  {Jiang}}]{Wang:2016rio}%
  \BibitemOpen
  \bibfield  {author} {\bibinfo {author} {\bibfnamefont {Lan}\ \bibnamefont
  {Wang}}, \bibinfo {author} {\bibfnamefont {Violeta}\ \bibnamefont
  {Gonzalez-Perez}}, \bibinfo {author} {\bibfnamefont {Lizhi}\ \bibnamefont
  {Xie}}, \bibinfo {author} {\bibfnamefont {Andrew~P.}\ \bibnamefont {Cooper}},
  \bibinfo {author} {\bibfnamefont {Carlos~S.}\ \bibnamefont {Frenk}}, \bibinfo
  {author} {\bibfnamefont {Liang}\ \bibnamefont {Gao}}, \bibinfo {author}
  {\bibfnamefont {Wojciech~A.}\ \bibnamefont {Hellwing}}, \bibinfo {author}
  {\bibfnamefont {John}\ \bibnamefont {Helly}}, \bibinfo {author}
  {\bibfnamefont {Mark~R.}\ \bibnamefont {Lovell}}, \ and\ \bibinfo {author}
  {\bibfnamefont {Lilian}\ \bibnamefont {Jiang}},\ }\bibfield  {title}
  {\enquote {\bibinfo {title} {{The galaxy population in cold and warm dark
  matter cosmologies}},}\ }\href {\doibase 10.1093/mnras/stx788} {\bibfield
  {journal} {\bibinfo  {journal} {Mon. Not. Roy. Astron. Soc.}\ }\textbf
  {\bibinfo {volume} {468}},\ \bibinfo {pages} {4579--4591} (\bibinfo {year}
  {2017})},\ \Eprint {http://arxiv.org/abs/1612.04540} {arXiv:1612.04540
  [astro-ph.GA]} \BibitemShut {NoStop}%
\bibitem [{\citenamefont {Bose}\ \emph {et~al.}(2016)\citenamefont {Bose},
  \citenamefont {Frenk}, \citenamefont {Hou}, \citenamefont {Lacey},\ and\
  \citenamefont {Lovell}}]{Bose:2016hlz}%
  \BibitemOpen
  \bibfield  {author} {\bibinfo {author} {\bibfnamefont {Sownak}\ \bibnamefont
  {Bose}}, \bibinfo {author} {\bibfnamefont {Carlos~S.}\ \bibnamefont {Frenk}},
  \bibinfo {author} {\bibfnamefont {Jun}\ \bibnamefont {Hou}}, \bibinfo
  {author} {\bibfnamefont {Cedric~G.}\ \bibnamefont {Lacey}}, \ and\ \bibinfo
  {author} {\bibfnamefont {Mark~R.}\ \bibnamefont {Lovell}},\ }\bibfield
  {title} {\enquote {\bibinfo {title} {{Reionization in sterile neutrino
  cosmologies}},}\ }\href {\doibase 10.1093/mnras/stw2288} {\bibfield
  {journal} {\bibinfo  {journal} {Mon. Not. Roy. Astron. Soc.}\ }\textbf
  {\bibinfo {volume} {463}},\ \bibinfo {pages} {3848--3859} (\bibinfo {year}
  {2016})},\ \Eprint {http://arxiv.org/abs/1605.03179} {arXiv:1605.03179
  [astro-ph.CO]} \BibitemShut {NoStop}%
\bibitem [{\citenamefont {Ross}\ \emph {et~al.}(2019)\citenamefont {Ross},
  \citenamefont {Dixon}, \citenamefont {Ghara}, \citenamefont {Iliev},\ and\
  \citenamefont {Mellema}}]{Ross:2018uhh}%
  \BibitemOpen
  \bibfield  {author} {\bibinfo {author} {\bibfnamefont {Hannah~E.}\
  \bibnamefont {Ross}}, \bibinfo {author} {\bibfnamefont {Keri~L.}\
  \bibnamefont {Dixon}}, \bibinfo {author} {\bibfnamefont {Raghunath}\
  \bibnamefont {Ghara}}, \bibinfo {author} {\bibfnamefont {Ilian~T.}\
  \bibnamefont {Iliev}}, \ and\ \bibinfo {author} {\bibfnamefont {Garrelt}\
  \bibnamefont {Mellema}},\ }\bibfield  {title} {\enquote {\bibinfo {title}
  {{Evaluating the QSO contribution to the 21-cm signal from the Cosmic
  Dawn}},}\ }\href {\doibase 10.1093/mnras/stz1220} {\bibfield  {journal}
  {\bibinfo  {journal} {Mon. Not. Roy. Astron. Soc.}\ }\textbf {\bibinfo
  {volume} {487}},\ \bibinfo {pages} {1101--1119} (\bibinfo {year} {2019})},\
  \Eprint {http://arxiv.org/abs/1808.03287} {arXiv:1808.03287 [astro-ph.CO]}
  \BibitemShut {NoStop}%
\bibitem [{\citenamefont {Safarzadeh}\ \emph {et~al.}(2018)\citenamefont
  {Safarzadeh}, \citenamefont {Scannapieco},\ and\ \citenamefont
  {Babul}}]{Safarzadeh:2018hhg}%
  \BibitemOpen
  \bibfield  {author} {\bibinfo {author} {\bibfnamefont {Mohammadtaher}\
  \bibnamefont {Safarzadeh}}, \bibinfo {author} {\bibfnamefont {Evan}\
  \bibnamefont {Scannapieco}}, \ and\ \bibinfo {author} {\bibfnamefont {Arif}\
  \bibnamefont {Babul}},\ }\bibfield  {title} {\enquote {\bibinfo {title} {{A
  limit on the warm dark matter particle mass from the redshifted 21 cm
  absorption line}},}\ }\href {\doibase 10.3847/2041-8213/aac5e0} {\bibfield
  {journal} {\bibinfo  {journal} {Astrophys. J.}\ }\textbf {\bibinfo {volume}
  {859}},\ \bibinfo {pages} {L18} (\bibinfo {year} {2018})},\ \Eprint
  {http://arxiv.org/abs/1803.08039} {arXiv:1803.08039 [astro-ph.CO]}
  \BibitemShut {NoStop}%
\bibitem [{\citenamefont {Bryan}\ and\ \citenamefont
  {Norman}(1998)}]{Bryan:1997dn}%
  \BibitemOpen
  \bibfield  {author} {\bibinfo {author} {\bibfnamefont {G.~L.}\ \bibnamefont
  {Bryan}}\ and\ \bibinfo {author} {\bibfnamefont {M.~L.}\ \bibnamefont
  {Norman}},\ }\bibfield  {title} {\enquote {\bibinfo {title} {{Statistical
  properties of x-ray clusters: Analytic and numerical comparisons}},}\ }\href
  {\doibase 10.1086/305262} {\bibfield  {journal} {\bibinfo  {journal}
  {Astrophys. J.}\ }\textbf {\bibinfo {volume} {495}},\ \bibinfo {pages} {80}
  (\bibinfo {year} {1998})},\ \Eprint {http://arxiv.org/abs/astro-ph/9710107}
  {arXiv:astro-ph/9710107 [astro-ph]} \BibitemShut {NoStop}%
\bibitem [{\citenamefont {{Oesch}}\ \emph {et~al.}(2018)\citenamefont
  {{Oesch}}, \citenamefont {{Bouwens}}, \citenamefont {{Illingworth}},
  \citenamefont {{Labb{\'e}}},\ and\ \citenamefont {{Stefanon}}}]{Oesch:17}%
  \BibitemOpen
  \bibfield  {author} {\bibinfo {author} {\bibfnamefont {P.~A.}\ \bibnamefont
  {{Oesch}}}, \bibinfo {author} {\bibfnamefont {R.~J.}\ \bibnamefont
  {{Bouwens}}}, \bibinfo {author} {\bibfnamefont {G.~D.}\ \bibnamefont
  {{Illingworth}}}, \bibinfo {author} {\bibfnamefont {I.}~\bibnamefont
  {{Labb{\'e}}}}, \ and\ \bibinfo {author} {\bibfnamefont {M.}~\bibnamefont
  {{Stefanon}}},\ }\bibfield  {title} {\enquote {\bibinfo {title} {{The Dearth
  of z {$\sim$} 10 Galaxies in All HST Legacy Fields -- The Rapid Evolution of
  the Galaxy Population in the First 500 Myr}},}\ }\href {\doibase
  10.3847/1538-4357/aab03f} {\bibfield  {journal} {\bibinfo  {journal} {\apj}\
  }\textbf {\bibinfo {volume} {855}},\ \bibinfo {eid} {105} (\bibinfo {year}
  {2018})},\ \Eprint {http://arxiv.org/abs/1710.11131} {arXiv:1710.11131}
  \BibitemShut {NoStop}%
\bibitem [{\citenamefont {{Dayal}}\ \emph {et~al.}(2014)\citenamefont
  {{Dayal}}, \citenamefont {{Ferrara}}, \citenamefont {{Dunlop}},\ and\
  \citenamefont {{Pacucci}}}]{Dayal:14}%
  \BibitemOpen
  \bibfield  {author} {\bibinfo {author} {\bibfnamefont {Pratika}\ \bibnamefont
  {{Dayal}}}, \bibinfo {author} {\bibfnamefont {Andrea}\ \bibnamefont
  {{Ferrara}}}, \bibinfo {author} {\bibfnamefont {James~S.}\ \bibnamefont
  {{Dunlop}}}, \ and\ \bibinfo {author} {\bibfnamefont {Fabio}\ \bibnamefont
  {{Pacucci}}},\ }\bibfield  {title} {\enquote {\bibinfo {title} {{Essential
  physics of early galaxy formation}},}\ }\href {\doibase
  10.1093/mnras/stu1848} {\bibfield  {journal} {\bibinfo  {journal} {Monthly
  Notices of the Royal Astronomical Society}\ }\textbf {\bibinfo {volume}
  {445}},\ \bibinfo {pages} {2545--2557} (\bibinfo {year} {2014})},\ \Eprint
  {http://arxiv.org/abs/1405.4862} {arXiv:1405.4862 [astro-ph.GA]} \BibitemShut
  {NoStop}%
\bibitem [{\citenamefont {{Sun}}\ and\ \citenamefont
  {{Furlanetto}}(2016)}]{Sun:15}%
  \BibitemOpen
  \bibfield  {author} {\bibinfo {author} {\bibfnamefont {G.}~\bibnamefont
  {{Sun}}}\ and\ \bibinfo {author} {\bibfnamefont {S.~R.}\ \bibnamefont
  {{Furlanetto}}},\ }\bibfield  {title} {\enquote {\bibinfo {title}
  {{Constraints on the star formation efficiency of galaxies during the epoch
  of reionization}},}\ }\href {\doibase 10.1093/mnras/stw980} {\bibfield
  {journal} {\bibinfo  {journal} {\mnras}\ }\textbf {\bibinfo {volume} {460}},\
  \bibinfo {pages} {417--433} (\bibinfo {year} {2016})},\ \Eprint
  {http://arxiv.org/abs/1512.06219} {arXiv:1512.06219} \BibitemShut {NoStop}%
\bibitem [{\citenamefont {{Mirocha}}\ \emph {et~al.}(2017)\citenamefont
  {{Mirocha}}, \citenamefont {{Furlanetto}},\ and\ \citenamefont
  {{Sun}}}]{Mirocha:16a}%
  \BibitemOpen
  \bibfield  {author} {\bibinfo {author} {\bibfnamefont {J.}~\bibnamefont
  {{Mirocha}}}, \bibinfo {author} {\bibfnamefont {S.~R.}\ \bibnamefont
  {{Furlanetto}}}, \ and\ \bibinfo {author} {\bibfnamefont {G.}~\bibnamefont
  {{Sun}}},\ }\bibfield  {title} {\enquote {\bibinfo {title} {{The global 21-cm
  signal in the context of the high- z galaxy luminosity function}},}\ }\href
  {\doibase 10.1093/mnras/stw2412} {\bibfield  {journal} {\bibinfo  {journal}
  {\mnras}\ }\textbf {\bibinfo {volume} {464}},\ \bibinfo {pages} {1365--1379}
  (\bibinfo {year} {2017})},\ \Eprint {http://arxiv.org/abs/1607.00386}
  {arXiv:1607.00386} \BibitemShut {NoStop}%
\bibitem [{\citenamefont {Mirocha}\ and\ \citenamefont
  {Furlanetto}(2019)}]{Mirocha:2018cih}%
  \BibitemOpen
  \bibfield  {author} {\bibinfo {author} {\bibfnamefont {Jordan}\ \bibnamefont
  {Mirocha}}\ and\ \bibinfo {author} {\bibfnamefont {Steven~R.}\ \bibnamefont
  {Furlanetto}},\ }\bibfield  {title} {\enquote {\bibinfo {title} {{What does
  the first highly-redshifted 21-cm detection tell us about early galaxies?}}}\
  }\href {\doibase 10.1093/mnras/sty3260} {\bibfield  {journal} {\bibinfo
  {journal} {Mon. Not. Roy. Astron. Soc.}\ }\textbf {\bibinfo {volume} {483}},\
  \bibinfo {pages} {1980--1992} (\bibinfo {year} {2019})},\ \Eprint
  {http://arxiv.org/abs/1803.03272} {arXiv:1803.03272 [astro-ph.GA]}
  \BibitemShut {NoStop}%
\bibitem [{\citenamefont {{Park}}\ \emph {et~al.}(2019)\citenamefont {{Park}},
  \citenamefont {{Mesinger}}, \citenamefont {{Greig}},\ and\ \citenamefont
  {{Gillet}}}]{Park:18}%
  \BibitemOpen
  \bibfield  {author} {\bibinfo {author} {\bibfnamefont {Jaehong}\ \bibnamefont
  {{Park}}}, \bibinfo {author} {\bibfnamefont {Andrei}\ \bibnamefont
  {{Mesinger}}}, \bibinfo {author} {\bibfnamefont {Bradley}\ \bibnamefont
  {{Greig}}}, \ and\ \bibinfo {author} {\bibfnamefont {Nicolas}\ \bibnamefont
  {{Gillet}}},\ }\bibfield  {title} {\enquote {\bibinfo {title} {{Inferring the
  astrophysics of reionization and cosmic dawn from galaxy luminosity functions
  and the 21-cm signal}},}\ }\href {\doibase 10.1093/mnras/stz032} {\bibfield
  {journal} {\bibinfo  {journal} {Monthly Notices of the Royal Astronomical
  Society}\ }\textbf {\bibinfo {volume} {484}},\ \bibinfo {pages} {933--949}
  (\bibinfo {year} {2019})},\ \Eprint {http://arxiv.org/abs/1809.08995}
  {arXiv:1809.08995 [astro-ph.GA]} \BibitemShut {NoStop}%
\bibitem [{\citenamefont {Sawala}\ \emph {et~al.}(2015)\citenamefont {Sawala}
  \emph {et~al.}}]{Sawala:2014baa}%
  \BibitemOpen
  \bibfield  {author} {\bibinfo {author} {\bibfnamefont {Till}\ \bibnamefont
  {Sawala}} \emph {et~al.},\ }\bibfield  {title} {\enquote {\bibinfo {title}
  {{Bent by baryons: the low mass galaxy-halo relation}},}\ }\href {\doibase
  10.1093/mnras/stu2753} {\bibfield  {journal} {\bibinfo  {journal} {Mon. Not.
  Roy. Astron. Soc.}\ }\textbf {\bibinfo {volume} {448}},\ \bibinfo {pages}
  {2941--2947} (\bibinfo {year} {2015})},\ \Eprint
  {http://arxiv.org/abs/1404.3724} {arXiv:1404.3724 [astro-ph.GA]} \BibitemShut
  {NoStop}%
\bibitem [{\citenamefont {Corasaniti}\ \emph {et~al.}(2017)\citenamefont
  {Corasaniti}, \citenamefont {Agarwal}, \citenamefont {Marsh},\ and\
  \citenamefont {Das}}]{Corasaniti:2016epp}%
  \BibitemOpen
  \bibfield  {author} {\bibinfo {author} {\bibfnamefont {P.~S.}\ \bibnamefont
  {Corasaniti}}, \bibinfo {author} {\bibfnamefont {S.}~\bibnamefont {Agarwal}},
  \bibinfo {author} {\bibfnamefont {D.~J.~E.}\ \bibnamefont {Marsh}}, \ and\
  \bibinfo {author} {\bibfnamefont {S.}~\bibnamefont {Das}},\ }\bibfield
  {title} {\enquote {\bibinfo {title} {{Constraints on dark matter scenarios
  from measurements of the galaxy luminosity function at high redshifts}},}\
  }\href {\doibase 10.1103/PhysRevD.95.083512} {\bibfield  {journal} {\bibinfo
  {journal} {Phys. Rev.}\ }\textbf {\bibinfo {volume} {D95}},\ \bibinfo {pages}
  {083512} (\bibinfo {year} {2017})},\ \Eprint
  {http://arxiv.org/abs/1611.05892} {arXiv:1611.05892 [astro-ph.CO]}
  \BibitemShut {NoStop}%
\bibitem [{\citenamefont {Menci}\ \emph {et~al.}(2018)\citenamefont {Menci},
  \citenamefont {Grazian}, \citenamefont {Lamastra}, \citenamefont {Calura},
  \citenamefont {Castellano},\ and\ \citenamefont {Santini}}]{Menci:2018lis}%
  \BibitemOpen
  \bibfield  {author} {\bibinfo {author} {\bibfnamefont {N.}~\bibnamefont
  {Menci}}, \bibinfo {author} {\bibfnamefont {A.}~\bibnamefont {Grazian}},
  \bibinfo {author} {\bibfnamefont {A.}~\bibnamefont {Lamastra}}, \bibinfo
  {author} {\bibfnamefont {F.}~\bibnamefont {Calura}}, \bibinfo {author}
  {\bibfnamefont {M.}~\bibnamefont {Castellano}}, \ and\ \bibinfo {author}
  {\bibfnamefont {P.}~\bibnamefont {Santini}},\ }\bibfield  {title} {\enquote
  {\bibinfo {title} {{Galaxy Formation in Sterile Neutrino Dark Matter
  Models}},}\ }\href {\doibase 10.3847/1538-4357/aaa773} {\bibfield  {journal}
  {\bibinfo  {journal} {Astrophys. J.}\ }\textbf {\bibinfo {volume} {854}},\
  \bibinfo {pages} {1} (\bibinfo {year} {2018})},\ \Eprint
  {http://arxiv.org/abs/1801.03697} {arXiv:1801.03697 [astro-ph.CO]}
  \BibitemShut {NoStop}%
\bibitem [{\citenamefont {Wise}\ \emph {et~al.}(2014)\citenamefont {Wise},
  \citenamefont {Demchenko}, \citenamefont {Halicek}, \citenamefont {Norman},
  \citenamefont {Turk}, \citenamefont {Abel},\ and\ \citenamefont
  {Smith}}]{Wise:2014vwa}%
  \BibitemOpen
  \bibfield  {author} {\bibinfo {author} {\bibfnamefont {John~H.}\ \bibnamefont
  {Wise}}, \bibinfo {author} {\bibfnamefont {Vasiliy~G.}\ \bibnamefont
  {Demchenko}}, \bibinfo {author} {\bibfnamefont {Martin~T.}\ \bibnamefont
  {Halicek}}, \bibinfo {author} {\bibfnamefont {Michael~L.}\ \bibnamefont
  {Norman}}, \bibinfo {author} {\bibfnamefont {Matthew~J.}\ \bibnamefont
  {Turk}}, \bibinfo {author} {\bibfnamefont {Tom}\ \bibnamefont {Abel}}, \ and\
  \bibinfo {author} {\bibfnamefont {Britton~D.}\ \bibnamefont {Smith}},\
  }\bibfield  {title} {\enquote {\bibinfo {title} {{The birth of a galaxy –
  III. Propelling reionization with the faintest galaxies}},}\ }\href {\doibase
  10.1093/mnras/stu979} {\bibfield  {journal} {\bibinfo  {journal} {Mon. Not.
  Roy. Astron. Soc.}\ }\textbf {\bibinfo {volume} {442}},\ \bibinfo {pages}
  {2560--2579} (\bibinfo {year} {2014})},\ \Eprint
  {http://arxiv.org/abs/1403.6123} {arXiv:1403.6123 [astro-ph.CO]} \BibitemShut
  {NoStop}%
\bibitem [{\citenamefont {{Xu}}\ \emph {et~al.}(2016)\citenamefont {{Xu}},
  \citenamefont {{Wise}}, \citenamefont {{Norman}}, \citenamefont {{Ahn}},\
  and\ \citenamefont {{O'Shea}}}]{Xu:16}%
  \BibitemOpen
  \bibfield  {author} {\bibinfo {author} {\bibfnamefont {H.}~\bibnamefont
  {{Xu}}}, \bibinfo {author} {\bibfnamefont {J.~H.}\ \bibnamefont {{Wise}}},
  \bibinfo {author} {\bibfnamefont {M.~L.}\ \bibnamefont {{Norman}}}, \bibinfo
  {author} {\bibfnamefont {K.}~\bibnamefont {{Ahn}}}, \ and\ \bibinfo {author}
  {\bibfnamefont {B.~W.}\ \bibnamefont {{O'Shea}}},\ }\bibfield  {title}
  {\enquote {\bibinfo {title} {{Galaxy Properties and UV Escape Fractions
  during the Epoch of Reionization: Results from the Renaissance
  Simulations}},}\ }\href {\doibase 10.3847/1538-4357/833/1/84} {\bibfield
  {journal} {\bibinfo  {journal} {\apj}\ }\textbf {\bibinfo {volume} {833}},\
  \bibinfo {eid} {84} (\bibinfo {year} {2016})},\ \Eprint
  {http://arxiv.org/abs/1604.07842} {arXiv:1604.07842} \BibitemShut {NoStop}%
\bibitem [{\citenamefont {Ma}\ \emph {et~al.}(2018)\citenamefont {Ma},
  \citenamefont {Hopkins}, \citenamefont {Garrison-Kimmel}, \citenamefont
  {Faucher-Giguère}, \citenamefont {Quataert}, \citenamefont {Boylan-Kolchin},
  \citenamefont {Hayward}, \citenamefont {Feldmann},\ and\ \citenamefont
  {Kereš}}]{Ma:2017avo}%
  \BibitemOpen
  \bibfield  {author} {\bibinfo {author} {\bibfnamefont {Xiangcheng}\
  \bibnamefont {Ma}}, \bibinfo {author} {\bibfnamefont {Philip~F.}\
  \bibnamefont {Hopkins}}, \bibinfo {author} {\bibfnamefont {Shea}\
  \bibnamefont {Garrison-Kimmel}}, \bibinfo {author} {\bibfnamefont
  {Claude-André}\ \bibnamefont {Faucher-Giguère}}, \bibinfo {author}
  {\bibfnamefont {Eliot}\ \bibnamefont {Quataert}}, \bibinfo {author}
  {\bibfnamefont {Michael}\ \bibnamefont {Boylan-Kolchin}}, \bibinfo {author}
  {\bibfnamefont {Christopher~C.}\ \bibnamefont {Hayward}}, \bibinfo {author}
  {\bibfnamefont {Robert}\ \bibnamefont {Feldmann}}, \ and\ \bibinfo {author}
  {\bibfnamefont {Dušan}\ \bibnamefont {Kereš}},\ }\bibfield  {title}
  {\enquote {\bibinfo {title} {{Simulating galaxies in the reionization era
  with FIRE-2: galaxy scaling relations, stellar mass functions, and luminosity
  functions}},}\ }\href {\doibase 10.1093/mnras/sty1024} {\bibfield  {journal}
  {\bibinfo  {journal} {Mon. Not. Roy. Astron. Soc.}\ }\textbf {\bibinfo
  {volume} {478}},\ \bibinfo {pages} {1694--1715} (\bibinfo {year} {2018})},\
  \Eprint {http://arxiv.org/abs/1706.06605} {arXiv:1706.06605 [astro-ph.GA]}
  \BibitemShut {NoStop}%
\bibitem [{\citenamefont {{Rosdahl}}\ \emph {et~al.}(2018)\citenamefont
  {{Rosdahl}}, \citenamefont {{Katz}}, \citenamefont {{Blaizot}}, \citenamefont
  {{Kimm}}, \citenamefont {{Michel-Dansac}}, \citenamefont {{Garel}},
  \citenamefont {{Haehnelt}}, \citenamefont {{Ocvirk}},\ and\ \citenamefont
  {{Teyssier}}}]{Rosdahl:18}%
  \BibitemOpen
  \bibfield  {author} {\bibinfo {author} {\bibfnamefont {J.}~\bibnamefont
  {{Rosdahl}}}, \bibinfo {author} {\bibfnamefont {H.}~\bibnamefont {{Katz}}},
  \bibinfo {author} {\bibfnamefont {J.}~\bibnamefont {{Blaizot}}}, \bibinfo
  {author} {\bibfnamefont {T.}~\bibnamefont {{Kimm}}}, \bibinfo {author}
  {\bibfnamefont {L.}~\bibnamefont {{Michel-Dansac}}}, \bibinfo {author}
  {\bibfnamefont {T.}~\bibnamefont {{Garel}}}, \bibinfo {author} {\bibfnamefont
  {M.}~\bibnamefont {{Haehnelt}}}, \bibinfo {author} {\bibfnamefont
  {P.}~\bibnamefont {{Ocvirk}}}, \ and\ \bibinfo {author} {\bibfnamefont
  {R.}~\bibnamefont {{Teyssier}}},\ }\bibfield  {title} {\enquote {\bibinfo
  {title} {{The SPHINX cosmological simulations of the first billion years: the
  impact of binary stars on reionization}},}\ }\href {\doibase
  10.1093/mnras/sty1655} {\bibfield  {journal} {\bibinfo  {journal} {\mnras}\
  }\textbf {\bibinfo {volume} {479}},\ \bibinfo {pages} {994--1016} (\bibinfo
  {year} {2018})},\ \Eprint {http://arxiv.org/abs/1801.07259}
  {arXiv:1801.07259} \BibitemShut {NoStop}%
\bibitem [{\citenamefont {Sharma}\ and\ \citenamefont
  {Theuns}(2019)}]{Sharma:2019mrx}%
  \BibitemOpen
  \bibfield  {author} {\bibinfo {author} {\bibfnamefont {Mahavir}\ \bibnamefont
  {Sharma}}\ and\ \bibinfo {author} {\bibfnamefont {Tom}\ \bibnamefont
  {Theuns}},\ }\bibfield  {title} {\enquote {\bibinfo {title} {{The
  $I\kappa\epsilon\alpha$ model of feedback-regulated galaxy formation}},}\
  }\href@noop {} {\  (\bibinfo {year} {2019})},\ \Eprint
  {http://arxiv.org/abs/1906.10135} {arXiv:1906.10135 [astro-ph.GA]}
  \BibitemShut {NoStop}%
\bibitem [{\citenamefont {Dayal}\ and\ \citenamefont
  {Ferrara}(2018)}]{Dayal:2018hft}%
  \BibitemOpen
  \bibfield  {author} {\bibinfo {author} {\bibfnamefont {Pratika}\ \bibnamefont
  {Dayal}}\ and\ \bibinfo {author} {\bibfnamefont {Andrea}\ \bibnamefont
  {Ferrara}},\ }\bibfield  {title} {\enquote {\bibinfo {title} {{Early galaxy
  formation and its large-scale effects}},}\ }\href {\doibase
  10.1016/j.physrep.2018.10.002} {\bibfield  {journal} {\bibinfo  {journal}
  {Phys. Rept.}\ }\textbf {\bibinfo {volume} {780-782}},\ \bibinfo {pages}
  {1--64} (\bibinfo {year} {2018})},\ \Eprint {http://arxiv.org/abs/1809.09136}
  {arXiv:1809.09136 [astro-ph.GA]} \BibitemShut {NoStop}%
\bibitem [{\citenamefont {Laine}\ and\ \citenamefont
  {Shaposhnikov}(2008)}]{Laine:2008pg}%
  \BibitemOpen
  \bibfield  {author} {\bibinfo {author} {\bibfnamefont {M.}~\bibnamefont
  {Laine}}\ and\ \bibinfo {author} {\bibfnamefont {M.}~\bibnamefont
  {Shaposhnikov}},\ }\bibfield  {title} {\enquote {\bibinfo {title} {{Sterile
  neutrino dark matter as a consequence of nuMSM-induced lepton asymmetry}},}\
  }\href {\doibase 10.1088/1475-7516/2008/06/031} {\bibfield  {journal}
  {\bibinfo  {journal} {JCAP}\ }\textbf {\bibinfo {volume} {0806}},\ \bibinfo
  {pages} {031} (\bibinfo {year} {2008})},\ \Eprint
  {http://arxiv.org/abs/0804.4543} {arXiv:0804.4543 [hep-ph]} \BibitemShut
  {NoStop}%
\bibitem [{\citenamefont {Boyarsky}\ \emph
  {et~al.}(2009{\natexlab{b}})\citenamefont {Boyarsky}, \citenamefont
  {Ruchayskiy},\ and\ \citenamefont {Shaposhnikov}}]{Boyarsky:2009ix}%
  \BibitemOpen
  \bibfield  {author} {\bibinfo {author} {\bibfnamefont {Alexey}\ \bibnamefont
  {Boyarsky}}, \bibinfo {author} {\bibfnamefont {Oleg}\ \bibnamefont
  {Ruchayskiy}}, \ and\ \bibinfo {author} {\bibfnamefont {Mikhail}\
  \bibnamefont {Shaposhnikov}},\ }\bibfield  {title} {\enquote {\bibinfo
  {title} {{The Role of sterile neutrinos in cosmology and astrophysics}},}\
  }\href {\doibase 10.1146/annurev.nucl.010909.083654} {\bibfield  {journal}
  {\bibinfo  {journal} {Ann. Rev. Nucl. Part. Sci.}\ }\textbf {\bibinfo
  {volume} {59}},\ \bibinfo {pages} {191--214} (\bibinfo {year}
  {2009}{\natexlab{b}})},\ \Eprint {http://arxiv.org/abs/0901.0011}
  {arXiv:0901.0011 [hep-ph]} \BibitemShut {NoStop}%
\bibitem [{\citenamefont {Bulbul}\ \emph {et~al.}(2014)\citenamefont {Bulbul},
  \citenamefont {Markevitch}, \citenamefont {Foster}, \citenamefont {Smith},
  \citenamefont {Loewenstein},\ and\ \citenamefont {Randall}}]{Bulbul:2014sua}%
  \BibitemOpen
  \bibfield  {author} {\bibinfo {author} {\bibfnamefont {Esra}\ \bibnamefont
  {Bulbul}}, \bibinfo {author} {\bibfnamefont {Maxim}\ \bibnamefont
  {Markevitch}}, \bibinfo {author} {\bibfnamefont {Adam}\ \bibnamefont
  {Foster}}, \bibinfo {author} {\bibfnamefont {Randall~K.}\ \bibnamefont
  {Smith}}, \bibinfo {author} {\bibfnamefont {Michael}\ \bibnamefont
  {Loewenstein}}, \ and\ \bibinfo {author} {\bibfnamefont {Scott~W.}\
  \bibnamefont {Randall}},\ }\bibfield  {title} {\enquote {\bibinfo {title}
  {{Detection of An Unidentified Emission Line in the Stacked X-ray spectrum of
  Galaxy Clusters}},}\ }\href {\doibase 10.1088/0004-637X/789/1/13} {\bibfield
  {journal} {\bibinfo  {journal} {Astrophys. J.}\ }\textbf {\bibinfo {volume}
  {789}},\ \bibinfo {pages} {13} (\bibinfo {year} {2014})},\ \Eprint
  {http://arxiv.org/abs/1402.2301} {arXiv:1402.2301 [astro-ph.CO]} \BibitemShut
  {NoStop}%
\bibitem [{\citenamefont {Boyarsky}\ \emph {et~al.}(2014)\citenamefont
  {Boyarsky}, \citenamefont {Ruchayskiy}, \citenamefont {Iakubovskyi},\ and\
  \citenamefont {Franse}}]{Boyarsky:2014jta}%
  \BibitemOpen
  \bibfield  {author} {\bibinfo {author} {\bibfnamefont {Alexey}\ \bibnamefont
  {Boyarsky}}, \bibinfo {author} {\bibfnamefont {Oleg}\ \bibnamefont
  {Ruchayskiy}}, \bibinfo {author} {\bibfnamefont {Dmytro}\ \bibnamefont
  {Iakubovskyi}}, \ and\ \bibinfo {author} {\bibfnamefont {Jeroen}\
  \bibnamefont {Franse}},\ }\bibfield  {title} {\enquote {\bibinfo {title}
  {{Unidentified Line in X-Ray Spectra of the Andromeda Galaxy and Perseus
  Galaxy Cluster}},}\ }\href {\doibase 10.1103/PhysRevLett.113.251301}
  {\bibfield  {journal} {\bibinfo  {journal} {Phys. Rev. Lett.}\ }\textbf
  {\bibinfo {volume} {113}},\ \bibinfo {pages} {251301} (\bibinfo {year}
  {2014})},\ \Eprint {http://arxiv.org/abs/1402.4119} {arXiv:1402.4119
  [astro-ph.CO]} \BibitemShut {NoStop}%
\bibitem [{\citenamefont {Boyarsky}\ \emph {et~al.}(2015)\citenamefont
  {Boyarsky}, \citenamefont {Franse}, \citenamefont {Iakubovskyi},\ and\
  \citenamefont {Ruchayskiy}}]{Boyarsky:2014ska}%
  \BibitemOpen
  \bibfield  {author} {\bibinfo {author} {\bibfnamefont {Alexey}\ \bibnamefont
  {Boyarsky}}, \bibinfo {author} {\bibfnamefont {Jeroen}\ \bibnamefont
  {Franse}}, \bibinfo {author} {\bibfnamefont {Dmytro}\ \bibnamefont
  {Iakubovskyi}}, \ and\ \bibinfo {author} {\bibfnamefont {Oleg}\ \bibnamefont
  {Ruchayskiy}},\ }\bibfield  {title} {\enquote {\bibinfo {title} {{Checking
  the Dark Matter Origin of a 3.53 keV Line with the Milky Way Center}},}\
  }\href {\doibase 10.1103/PhysRevLett.115.161301} {\bibfield  {journal}
  {\bibinfo  {journal} {Phys. Rev. Lett.}\ }\textbf {\bibinfo {volume} {115}},\
  \bibinfo {pages} {161301} (\bibinfo {year} {2015})},\ \Eprint
  {http://arxiv.org/abs/1408.2503} {arXiv:1408.2503 [astro-ph.CO]} \BibitemShut
  {NoStop}%
\bibitem [{\citenamefont {Iakubovskyi}\ \emph {et~al.}(2015)\citenamefont
  {Iakubovskyi}, \citenamefont {Bulbul}, \citenamefont {Foster}, \citenamefont
  {Savchenko},\ and\ \citenamefont {Sadova}}]{Iakubovskyi:2015dna}%
  \BibitemOpen
  \bibfield  {author} {\bibinfo {author} {\bibfnamefont {Dmytro}\ \bibnamefont
  {Iakubovskyi}}, \bibinfo {author} {\bibfnamefont {Esra}\ \bibnamefont
  {Bulbul}}, \bibinfo {author} {\bibfnamefont {Adam~R.}\ \bibnamefont
  {Foster}}, \bibinfo {author} {\bibfnamefont {Denys}\ \bibnamefont
  {Savchenko}}, \ and\ \bibinfo {author} {\bibfnamefont {Valentyna}\
  \bibnamefont {Sadova}},\ }\bibfield  {title} {\enquote {\bibinfo {title}
  {{Testing the origin of ~3.55 keV line in individual galaxy clusters observed
  with XMM-Newton}},}\ }\href@noop {} {\  (\bibinfo {year} {2015})},\ \Eprint
  {http://arxiv.org/abs/1508.05186} {arXiv:1508.05186 [astro-ph.HE]}
  \BibitemShut {NoStop}%
\bibitem [{\citenamefont {Ruchayskiy}\ \emph {et~al.}(2016)\citenamefont
  {Ruchayskiy}, \citenamefont {Boyarsky}, \citenamefont {Iakubovskyi},
  \citenamefont {Bulbul}, \citenamefont {Eckert}, \citenamefont {Franse},
  \citenamefont {Malyshev}, \citenamefont {Markevitch},\ and\ \citenamefont
  {Neronov}}]{Ruchayskiy:2015onc}%
  \BibitemOpen
  \bibfield  {author} {\bibinfo {author} {\bibfnamefont {Oleg}\ \bibnamefont
  {Ruchayskiy}}, \bibinfo {author} {\bibfnamefont {Alexey}\ \bibnamefont
  {Boyarsky}}, \bibinfo {author} {\bibfnamefont {Dmytro}\ \bibnamefont
  {Iakubovskyi}}, \bibinfo {author} {\bibfnamefont {Esra}\ \bibnamefont
  {Bulbul}}, \bibinfo {author} {\bibfnamefont {Dominique}\ \bibnamefont
  {Eckert}}, \bibinfo {author} {\bibfnamefont {Jeroen}\ \bibnamefont {Franse}},
  \bibinfo {author} {\bibfnamefont {Denys}\ \bibnamefont {Malyshev}}, \bibinfo
  {author} {\bibfnamefont {Maxim}\ \bibnamefont {Markevitch}}, \ and\ \bibinfo
  {author} {\bibfnamefont {Andrii}\ \bibnamefont {Neronov}},\ }\bibfield
  {title} {\enquote {\bibinfo {title} {{Searching for decaying dark matter in
  deep XMM–Newton observation of the Draco dwarf spheroidal}},}\ }\href
  {\doibase 10.1093/mnras/stw1026} {\bibfield  {journal} {\bibinfo  {journal}
  {Mon. Not. Roy. Astron. Soc.}\ }\textbf {\bibinfo {volume} {460}},\ \bibinfo
  {pages} {1390--1398} (\bibinfo {year} {2016})},\ \Eprint
  {http://arxiv.org/abs/1512.07217} {arXiv:1512.07217 [astro-ph.HE]}
  \BibitemShut {NoStop}%
\bibitem [{\citenamefont {Franse}\ \emph {et~al.}(2016)\citenamefont {Franse}
  \emph {et~al.}}]{Franse:2016dln}%
  \BibitemOpen
  \bibfield  {author} {\bibinfo {author} {\bibfnamefont {Jeroen}\ \bibnamefont
  {Franse}} \emph {et~al.},\ }\bibfield  {title} {\enquote {\bibinfo {title}
  {{Radial Profile of the 3.55 keV line out to $R_{200}$ in the Perseus
  Cluster}},}\ }\href {\doibase 10.3847/0004-637X/829/2/124} {\bibfield
  {journal} {\bibinfo  {journal} {Astrophys. J.}\ }\textbf {\bibinfo {volume}
  {829}},\ \bibinfo {pages} {124} (\bibinfo {year} {2016})},\ \Eprint
  {http://arxiv.org/abs/1604.01759} {arXiv:1604.01759 [astro-ph.CO]}
  \BibitemShut {NoStop}%
\bibitem [{\citenamefont {Drewes}\ \emph {et~al.}(2017)\citenamefont {Drewes}
  \emph {et~al.}}]{Adhikari:2016bei}%
  \BibitemOpen
  \bibfield  {author} {\bibinfo {author} {\bibfnamefont {M.}~\bibnamefont
  {Drewes}} \emph {et~al.},\ }\bibfield  {title} {\enquote {\bibinfo {title}
  {{A White Paper on keV Sterile Neutrino Dark Matter}},}\ }\href {\doibase
  10.1088/1475-7516/2017/01/025} {\bibfield  {journal} {\bibinfo  {journal}
  {JCAP}\ }\textbf {\bibinfo {volume} {1701}},\ \bibinfo {pages} {025}
  (\bibinfo {year} {2017})},\ \Eprint {http://arxiv.org/abs/1602.04816}
  {arXiv:1602.04816 [hep-ph]} \BibitemShut {NoStop}%
\bibitem [{\citenamefont {Abazajian}(2017)}]{Abazajian:2017tcc}%
  \BibitemOpen
  \bibfield  {author} {\bibinfo {author} {\bibfnamefont {Kevork~N.}\
  \bibnamefont {Abazajian}},\ }\bibfield  {title} {\enquote {\bibinfo {title}
  {{Sterile neutrinos in cosmology}},}\ }\href {\doibase
  10.1016/j.physrep.2017.10.003} {\bibfield  {journal} {\bibinfo  {journal}
  {Phys. Rept.}\ }\textbf {\bibinfo {volume} {711-712}},\ \bibinfo {pages}
  {1--28} (\bibinfo {year} {2017})},\ \Eprint {http://arxiv.org/abs/1705.01837}
  {arXiv:1705.01837 [hep-ph]} \BibitemShut {NoStop}%
\bibitem [{\citenamefont {Garzilli}\ \emph {et~al.}(2017)\citenamefont
  {Garzilli}, \citenamefont {Boyarsky},\ and\ \citenamefont
  {Ruchayskiy}}]{Garzilli:2015iwa}%
  \BibitemOpen
  \bibfield  {author} {\bibinfo {author} {\bibfnamefont {Antonella}\
  \bibnamefont {Garzilli}}, \bibinfo {author} {\bibfnamefont {Alexey}\
  \bibnamefont {Boyarsky}}, \ and\ \bibinfo {author} {\bibfnamefont {Oleg}\
  \bibnamefont {Ruchayskiy}},\ }\bibfield  {title} {\enquote {\bibinfo {title}
  {{Cutoff in the Lyman-alpha forest power spectrum: warm IGM or warm dark
  matter?}}}\ }\href {\doibase 10.1016/j.physletb.2017.08.022} {\bibfield
  {journal} {\bibinfo  {journal} {Phys. Lett.}\ }\textbf {\bibinfo {volume}
  {B773}},\ \bibinfo {pages} {258--264} (\bibinfo {year} {2017})},\ \Eprint
  {http://arxiv.org/abs/1510.07006} {arXiv:1510.07006 [astro-ph.CO]}
  \BibitemShut {NoStop}%
\bibitem [{\citenamefont {Baur}\ \emph {et~al.}(2017)\citenamefont {Baur},
  \citenamefont {Palanque-Delabrouille}, \citenamefont {Yeche}, \citenamefont
  {Boyarsky}, \citenamefont {Ruchayskiy}, \citenamefont {Armengaud},\ and\
  \citenamefont {Lesgourgues}}]{Baur:2017stq}%
  \BibitemOpen
  \bibfield  {author} {\bibinfo {author} {\bibfnamefont {Julien}\ \bibnamefont
  {Baur}}, \bibinfo {author} {\bibfnamefont {Nathalie}\ \bibnamefont
  {Palanque-Delabrouille}}, \bibinfo {author} {\bibfnamefont {Christophe}\
  \bibnamefont {Yeche}}, \bibinfo {author} {\bibfnamefont {Alexey}\
  \bibnamefont {Boyarsky}}, \bibinfo {author} {\bibfnamefont {Oleg}\
  \bibnamefont {Ruchayskiy}}, \bibinfo {author} {\bibfnamefont {Éric}\
  \bibnamefont {Armengaud}}, \ and\ \bibinfo {author} {\bibfnamefont {Julien}\
  \bibnamefont {Lesgourgues}},\ }\bibfield  {title} {\enquote {\bibinfo {title}
  {{Constraints from Ly-$\alpha$ forests on non-thermal dark matter including
  resonantly-produced sterile neutrinos}},}\ }\href {\doibase
  10.1088/1475-7516/2017/12/013} {\bibfield  {journal} {\bibinfo  {journal}
  {JCAP}\ }\textbf {\bibinfo {volume} {1712}},\ \bibinfo {pages} {013}
  (\bibinfo {year} {2017})},\ \Eprint {http://arxiv.org/abs/1706.03118}
  {arXiv:1706.03118 [astro-ph.CO]} \BibitemShut {NoStop}%
\bibitem [{\citenamefont {Garzilli}\ \emph {et~al.}(2019)\citenamefont
  {Garzilli}, \citenamefont {Magalich}, \citenamefont {Theuns}, \citenamefont
  {Frenk}, \citenamefont {Weniger}, \citenamefont {Ruchayskiy},\ and\
  \citenamefont {Boyarsky}}]{Garzilli:2018jqh}%
  \BibitemOpen
  \bibfield  {author} {\bibinfo {author} {\bibfnamefont {Antonella}\
  \bibnamefont {Garzilli}}, \bibinfo {author} {\bibfnamefont {Andrii}\
  \bibnamefont {Magalich}}, \bibinfo {author} {\bibfnamefont {Tom}\
  \bibnamefont {Theuns}}, \bibinfo {author} {\bibfnamefont {Carlos~S.}\
  \bibnamefont {Frenk}}, \bibinfo {author} {\bibfnamefont {Christoph}\
  \bibnamefont {Weniger}}, \bibinfo {author} {\bibfnamefont {Oleg}\
  \bibnamefont {Ruchayskiy}}, \ and\ \bibinfo {author} {\bibfnamefont {Alexey}\
  \bibnamefont {Boyarsky}},\ }\bibfield  {title} {\enquote {\bibinfo {title}
  {{The Lyman-$\alpha$ forest as a diagnostic of the nature of the dark
  matter}},}\ }\href {\doibase 10.1093/mnras/stz2188} {\bibfield  {journal}
  {\bibinfo  {journal} {Mon. Not. Roy. Astron. Soc.}\ }\textbf {\bibinfo
  {volume} {489}},\ \bibinfo {pages} {3456--3471} (\bibinfo {year} {2019})},\
  \Eprint {http://arxiv.org/abs/1809.06585} {arXiv:1809.06585 [astro-ph.CO]}
  \BibitemShut {NoStop}%
\bibitem [{\citenamefont {Rudakovskiy}\ and\ \citenamefont
  {Iakubovskyi}(2016)}]{Rudakovskiy:2016ngi}%
  \BibitemOpen
  \bibfield  {author} {\bibinfo {author} {\bibfnamefont {Anton}\ \bibnamefont
  {Rudakovskiy}}\ and\ \bibinfo {author} {\bibfnamefont {Dmytro}\ \bibnamefont
  {Iakubovskyi}},\ }\bibfield  {title} {\enquote {\bibinfo {title} {{Influence
  of ~7 keV sterile neutrino dark matter on the process of reionization}},}\
  }\href {\doibase 10.1088/1475-7516/2016/06/017} {\bibfield  {journal}
  {\bibinfo  {journal} {JCAP}\ }\textbf {\bibinfo {volume} {1606}},\ \bibinfo
  {pages} {017} (\bibinfo {year} {2016})},\ \Eprint
  {http://arxiv.org/abs/1604.01341} {arXiv:1604.01341 [astro-ph.CO]}
  \BibitemShut {NoStop}%
\bibitem [{\citenamefont {Rudakovskyi}\ and\ \citenamefont
  {Iakubovskyi}(2019)}]{Rudakovskyi:2018jfc}%
  \BibitemOpen
  \bibfield  {author} {\bibinfo {author} {\bibfnamefont {Anton}\ \bibnamefont
  {Rudakovskyi}}\ and\ \bibinfo {author} {\bibfnamefont {Dmytro}\ \bibnamefont
  {Iakubovskyi}},\ }\bibfield  {title} {\enquote {\bibinfo {title} {{Dark
  matter model favoured by reionization data: 7 keV sterile neutrino versus
  cold dark matter}},}\ }\href {\doibase 10.1093/mnras/sty3057} {\bibfield
  {journal} {\bibinfo  {journal} {Mon. Not. Roy. Astron. Soc.}\ }\textbf
  {\bibinfo {volume} {483}},\ \bibinfo {pages} {4080--4084} (\bibinfo {year}
  {2019})},\ \Eprint {http://arxiv.org/abs/1811.02799} {arXiv:1811.02799
  [astro-ph.CO]} \BibitemShut {NoStop}%
\bibitem [{\citenamefont {Menci}\ \emph {et~al.}(2016)\citenamefont {Menci},
  \citenamefont {Grazian}, \citenamefont {Castellano},\ and\ \citenamefont
  {Sanchez}}]{Menci:2016eui}%
  \BibitemOpen
  \bibfield  {author} {\bibinfo {author} {\bibfnamefont {N.}~\bibnamefont
  {Menci}}, \bibinfo {author} {\bibfnamefont {A.}~\bibnamefont {Grazian}},
  \bibinfo {author} {\bibfnamefont {M.}~\bibnamefont {Castellano}}, \ and\
  \bibinfo {author} {\bibfnamefont {N.~G.}\ \bibnamefont {Sanchez}},\
  }\bibfield  {title} {\enquote {\bibinfo {title} {{A Stringent Limit on the
  Warm Dark Matter Particle Masses from the Abundance of z=6 Galaxies in the
  Hubble Frontier Fields}},}\ }\href {\doibase 10.3847/2041-8205/825/1/L1}
  {\bibfield  {journal} {\bibinfo  {journal} {Astrophys. J.}\ }\textbf
  {\bibinfo {volume} {825}},\ \bibinfo {pages} {L1} (\bibinfo {year} {2016})},\
  \Eprint {http://arxiv.org/abs/1606.02530} {arXiv:1606.02530 [astro-ph.CO]}
  \BibitemShut {NoStop}%
\bibitem [{\citenamefont {Birrer}\ \emph {et~al.}(2017)\citenamefont {Birrer},
  \citenamefont {Amara},\ and\ \citenamefont {Refregier}}]{Birrer:2017rpp}%
  \BibitemOpen
  \bibfield  {author} {\bibinfo {author} {\bibfnamefont {Simon}\ \bibnamefont
  {Birrer}}, \bibinfo {author} {\bibfnamefont {Adam}\ \bibnamefont {Amara}}, \
  and\ \bibinfo {author} {\bibfnamefont {Alexandre}\ \bibnamefont
  {Refregier}},\ }\bibfield  {title} {\enquote {\bibinfo {title} {{Lensing
  substructure quantification in RXJ1131-1231: A 2 keV lower bound on dark
  matter thermal relic mass}},}\ }\href {\doibase
  10.1088/1475-7516/2017/05/037} {\bibfield  {journal} {\bibinfo  {journal}
  {JCAP}\ }\textbf {\bibinfo {volume} {1705}},\ \bibinfo {pages} {037}
  (\bibinfo {year} {2017})},\ \Eprint {http://arxiv.org/abs/1702.00009}
  {arXiv:1702.00009 [astro-ph.CO]} \BibitemShut {NoStop}%
\bibitem [{\citenamefont {Vegetti}\ \emph {et~al.}(2018)\citenamefont
  {Vegetti}, \citenamefont {Despali}, \citenamefont {Lovell},\ and\
  \citenamefont {Enzi}}]{Vegetti:2018dly}%
  \BibitemOpen
  \bibfield  {author} {\bibinfo {author} {\bibfnamefont {S.}~\bibnamefont
  {Vegetti}}, \bibinfo {author} {\bibfnamefont {G.}~\bibnamefont {Despali}},
  \bibinfo {author} {\bibfnamefont {M.~R.}\ \bibnamefont {Lovell}}, \ and\
  \bibinfo {author} {\bibfnamefont {W.}~\bibnamefont {Enzi}},\ }\bibfield
  {title} {\enquote {\bibinfo {title} {{Constraining sterile neutrino
  cosmologies with strong gravitational lensing observations at redshift z ∼
  0.2}},}\ }\href {\doibase 10.1093/mnras/sty2393} {\bibfield  {journal}
  {\bibinfo  {journal} {Mon. Not. Roy. Astron. Soc.}\ }\textbf {\bibinfo
  {volume} {481}},\ \bibinfo {pages} {3661--3669} (\bibinfo {year} {2018})},\
  \Eprint {http://arxiv.org/abs/1801.01505} {arXiv:1801.01505 [astro-ph.CO]}
  \BibitemShut {NoStop}%
\bibitem [{\citenamefont {Lovell}\ \emph {et~al.}(2014)\citenamefont {Lovell},
  \citenamefont {Frenk}, \citenamefont {Eke}, \citenamefont {Jenkins},
  \citenamefont {Gao},\ and\ \citenamefont {Theuns}}]{Lovell:2013ola}%
  \BibitemOpen
  \bibfield  {author} {\bibinfo {author} {\bibfnamefont {Mark~R.}\ \bibnamefont
  {Lovell}}, \bibinfo {author} {\bibfnamefont {Carlos~S.}\ \bibnamefont
  {Frenk}}, \bibinfo {author} {\bibfnamefont {Vincent~R.}\ \bibnamefont {Eke}},
  \bibinfo {author} {\bibfnamefont {Adrian}\ \bibnamefont {Jenkins}}, \bibinfo
  {author} {\bibfnamefont {Liang}\ \bibnamefont {Gao}}, \ and\ \bibinfo
  {author} {\bibfnamefont {Tom}\ \bibnamefont {Theuns}},\ }\bibfield  {title}
  {\enquote {\bibinfo {title} {{The properties of warm dark matter haloes}},}\
  }\href {\doibase 10.1093/mnras/stt2431} {\bibfield  {journal} {\bibinfo
  {journal} {Mon. Not. Roy. Astron. Soc.}\ }\textbf {\bibinfo {volume} {439}},\
  \bibinfo {pages} {300--317} (\bibinfo {year} {2014})},\ \Eprint
  {http://arxiv.org/abs/1308.1399} {arXiv:1308.1399 [astro-ph.CO]} \BibitemShut
  {NoStop}%
\bibitem [{\citenamefont {Kennedy}\ \emph {et~al.}(2014)\citenamefont
  {Kennedy}, \citenamefont {Frenk}, \citenamefont {Cole},\ and\ \citenamefont
  {Benson}}]{Kennedy:2013uta}%
  \BibitemOpen
  \bibfield  {author} {\bibinfo {author} {\bibfnamefont {Rachel}\ \bibnamefont
  {Kennedy}}, \bibinfo {author} {\bibfnamefont {Carlos}\ \bibnamefont {Frenk}},
  \bibinfo {author} {\bibfnamefont {Shaun}\ \bibnamefont {Cole}}, \ and\
  \bibinfo {author} {\bibfnamefont {Andrew}\ \bibnamefont {Benson}},\
  }\bibfield  {title} {\enquote {\bibinfo {title} {{Constraining the warm dark
  matter particle mass with Milky Way satellites}},}\ }\href {\doibase
  10.1093/mnras/stu719} {\bibfield  {journal} {\bibinfo  {journal} {Mon. Not.
  Roy. Astron. Soc.}\ }\textbf {\bibinfo {volume} {442}},\ \bibinfo {pages}
  {2487--2495} (\bibinfo {year} {2014})},\ \Eprint
  {http://arxiv.org/abs/1310.7739} {arXiv:1310.7739 [astro-ph.CO]} \BibitemShut
  {NoStop}%
\bibitem [{\citenamefont {Mesinger}\ \emph {et~al.}(2013)\citenamefont
  {Mesinger}, \citenamefont {Ferrara},\ and\ \citenamefont
  {Spiegel}}]{Mesinger:2012ys}%
  \BibitemOpen
  \bibfield  {author} {\bibinfo {author} {\bibfnamefont {Andrei}\ \bibnamefont
  {Mesinger}}, \bibinfo {author} {\bibfnamefont {Andrea}\ \bibnamefont
  {Ferrara}}, \ and\ \bibinfo {author} {\bibfnamefont {David~S.}\ \bibnamefont
  {Spiegel}},\ }\bibfield  {title} {\enquote {\bibinfo {title} {{Signatures of
  X-rays in the early Universe}},}\ }\href {\doibase 10.1093/mnras/stt198}
  {\bibfield  {journal} {\bibinfo  {journal} {Mon. Not. Roy. Astron. Soc.}\
  }\textbf {\bibinfo {volume} {431}},\ \bibinfo {pages} {621} (\bibinfo {year}
  {2013})},\ \Eprint {http://arxiv.org/abs/1210.7319} {arXiv:1210.7319
  [astro-ph.CO]} \BibitemShut {NoStop}%
\bibitem [{\citenamefont {Yue}\ and\ \citenamefont {Chen}(2012)}]{Yue:2012na}%
  \BibitemOpen
  \bibfield  {author} {\bibinfo {author} {\bibfnamefont {Bin}\ \bibnamefont
  {Yue}}\ and\ \bibinfo {author} {\bibfnamefont {Xuelei}\ \bibnamefont
  {Chen}},\ }\bibfield  {title} {\enquote {\bibinfo {title} {{Reionization in
  the Warm Dark Matter Model}},}\ }\href {\doibase 10.1088/0004-637X/747/2/127}
  {\bibfield  {journal} {\bibinfo  {journal} {Astrophys. J.}\ }\textbf
  {\bibinfo {volume} {747}},\ \bibinfo {pages} {127} (\bibinfo {year}
  {2012})},\ \Eprint {http://arxiv.org/abs/1201.3686} {arXiv:1201.3686
  [astro-ph.CO]} \BibitemShut {NoStop}%
\bibitem [{\citenamefont {Muñoz}\ and\ \citenamefont
  {Loeb}(2018)}]{Munoz:2018pzp}%
  \BibitemOpen
  \bibfield  {author} {\bibinfo {author} {\bibfnamefont {Julian~B.}\
  \bibnamefont {Muñoz}}\ and\ \bibinfo {author} {\bibfnamefont {Abraham}\
  \bibnamefont {Loeb}},\ }\bibfield  {title} {\enquote {\bibinfo {title} {{A
  small amount of mini-charged dark matter could cool the baryons in the early
  Universe}},}\ }\href {\doibase 10.1038/s41586-018-0151-x} {\bibfield
  {journal} {\bibinfo  {journal} {Nature}\ }\textbf {\bibinfo {volume} {557}},\
  \bibinfo {pages} {684} (\bibinfo {year} {2018})},\ \Eprint
  {http://arxiv.org/abs/1802.10094} {arXiv:1802.10094 [astro-ph.CO]}
  \BibitemShut {NoStop}%
\bibitem [{\citenamefont {Galli}\ and\ \citenamefont
  {Palla}(1998)}]{Galli:1998dh}%
  \BibitemOpen
  \bibfield  {author} {\bibinfo {author} {\bibfnamefont {Daniele}\ \bibnamefont
  {Galli}}\ and\ \bibinfo {author} {\bibfnamefont {Francesco}\ \bibnamefont
  {Palla}},\ }\bibfield  {title} {\enquote {\bibinfo {title} {{The Chemistry of
  the early universe}},}\ }\href@noop {} {\bibfield  {journal} {\bibinfo
  {journal} {Astron. Astrophys.}\ }\textbf {\bibinfo {volume} {335}},\ \bibinfo
  {pages} {403--420} (\bibinfo {year} {1998})},\ \Eprint
  {http://arxiv.org/abs/astro-ph/9803315} {arXiv:astro-ph/9803315 [astro-ph]}
  \BibitemShut {NoStop}%
\end{thebibliography}%

\end{document}